\documentclass[10pt,a4paper]{article}

\usepackage{amsmath,amsfonts,amssymb}
\usepackage{graphicx}
\usepackage[colorlinks=true, allcolors=blue]{hyperref}

\usepackage{geometry}
\usepackage{hhline}
\usepackage{subfigure}
\graphicspath{{Figures/}}
\usepackage{bm}
\usepackage{textcomp}
\usepackage[dvipsnames]{xcolor}
\usepackage{multirow} 
\usepackage{booktabs}

\def\eg{{\it e.g., }}

\newcommand{\revision}[1]{{\textcolor{black} {#1}}}

\usepackage[utf8]{inputenc}
\usepackage[english]{babel}
\usepackage{hyperref}
\usepackage{authblk}

\title{Experimental learning of a hyperelastic behavior with a Physics-Augmented Neural Network}
\author[]{C. Jailin\footnote{Corresponding Author: \href{mailto:clement.jailin@ens-paris-saclay.fr}{clement.jailin@ens-paris-saclay.fr}}}
\author[]{A. Benady}
\author[]{R. Legroux}
\author[]{E. Baranger}
\affil[]{Universit\'e Paris-Saclay, CentraleSupélec, ENS Paris-Saclay, CNRS, \\ LMPS -- Laboratoire de M\'ecanique Paris-Saclay, Gif-sur-Yvette, France}
\date{}

\begin{document} 

\maketitle

\begin{abstract}
\textbf{Background}: The recent development of Physics-Augmented Neural Networks (PANN) opens new opportunities for modeling material behaviors. These approaches have demonstrated their efficiency when trained on synthetic cases.
\textbf{Objective}: This study aims to demonstrate the effectiveness of training PANN using real experimental data \revision{for modeling} hyperelastic behavior.
\textbf{Methods}: The approach involved two uni-axial experiments equipped with digital image correlation and force sensors. The tests achieved axial deformations exceeding 200\% and presented non-linear responses. Twenty loading steps extracted from one experiment were used to train the PANN. The model architecture was optimized based on results from a validation dataset, utilizing equilibrium gap loss computed on six loading steps. \revision{Finally, 544 loading steps from the
first experiment and 80 steps from a second independent experiment} were used for testing purposes.
\textbf{Results}: The PANN model effectively captured the hyperelastic behavior across and beyond the training loads, showing superior performance compared to the standard Neo-Hookean model when assessed using various evaluation metrics.
\textbf{Conclusions}: Training PANN with experimental mechanical data shows promising results, outperforming traditional modeling approaches.
\end{abstract}


\section{INTRODUCTION}
Recent advancements in data-driven methods have been changing the way constitutive laws are modeled and characterized in material sciences. The rise of these approaches, based on data learning, is mainly due to two factors. First, the development of artificial intelligence applied to the mechanical sciences allows the use of agile and efficient models capable of representing complex behaviors~\cite{dornheim2024neural}. 
The second factor is the recent development of modern experimental mechanics, such as full-field measurements. This gives access to the collection of large quantities of experimental data that can be used for model identification~\cite{neggers2018big,roux2020optimal}. 

Common challenges in learning constitutive behavior, such as physical inconsistency, poor generalization, and data access, are addressed by combining machine learning models with physical information~\cite{Herrmann2024DeepLI}. Different strategies have emerged in this context.
First, methods such as "data-driven identification" (DDI) (or referred to as "model-free") techniques~\cite{leygue2018data, leygue2019non, stainier2019model}. These approaches enforce compatible and balanced mechanical states and identify a constitutive law from displacement and force measurements. 
A second category, known as "physics-informed" ~\cite{raissi2019physics,abueidda2023enhanced,masi2022multiscale},  involves embedding physical knowledge into the neural network loss function used for the training model. This method combines a discrepancy measure and a penalty for deviating from physical laws in the loss function. 
The third strategy, often named "Physics-Augmented Neural Network" emphasizes enforcing physical knowledge within the model's architecture itself. Various frameworks~\cite{fernandez2021anisotropic, klein2022polyconvex,as2022mechanics,linden2023neural,linka2023new}, propose the integration of physical properties, such as thermodynamic constraints, into the architecture.  
These \revision{latter} techniques generally use Input-Convex Neural Networks~\cite{amos2017input} (ICNN), which enforce the convexity of free energy. 
Other thermodynamically consistent architectures exist, such as \cite{hernandez2021structure}, in which a network is constructed to preserve the metriplectic structure of dissipative systems (in the form of the so-called General Equation for the Non-Equilibrium Reversible-Irreversible Coupling - GENERIC).
It can be highlighted that recent data-driven techniques bridge the gap between physics-augmented and DDI approaches~\cite{costecalde2023data}.

Initially, mechanical tests relied on simple measurements (e.g., a strain gauge and an extensometer). This situation significantly changed in the 2000s, with an increasing amount of data being regularly collected during experimental mechanics tests~\cite{neggers2018big}. This is mainly due to the use of quantitative image processing methods based on increasingly high-resolution images or volumes. Digital Image Correlation (DIC)~\cite{lucas1981iterative,sutton1983determination}, or Digital Volume Correlation (DVC)~\cite{bay2008methods,buljac2018digital} involves measuring a displacement field between a reference image taken, for example, at the beginning of the test and an image in a deformed configuration. Using global DIC~\cite{hild2012comparison}, the measured displacement field is expressed through a finite element mesh. In addition to regularizing the problem, this allows for smooth interfaces with finite element softwares to identify mechanical properties (\eg Finite Element Model Updating~\cite{mathieu2015estimation}). The richness of these measurements provides the kinematics data used to feed data-driven models.

While displacement fields, and hence deformation fields, are measurable, one of the challenges with data-based methods is the need to collect a large number of stress fields. Experimentally, as stress measurements are not directly accessible, they can be estimated through auxiliary measurements, such as force measurements, and on all the free surfaces. 
In practice, most tests are instrumented with simple and few force sensors (uni-axial or bi-axial measurement cells). Recent developments have enabled the measurement of complex multiaxial forces or force distributions~\cite{dassonville2020toward}. In this context, moving away from the supervised training framework, which depends on known stress labels, is crucial for developing truly efficient and practically applicable data-driven constitutive models~\cite{dalemat2019measuring}.

To overcome these limitations, a framework called \emph{Efficient Unsupervised Constitutive Law Identification and Discovery} (EUCLID)~\cite{flaschel2021unsupervised} was proposed. The method is termed \emph{unsupervised}, in the sense that it does not require stress data but only global reaction forces and full-field displacement data, accessible through full-field measurements. The EUCLID method includes a representation of the constitutive law that aims to map the displacement measurement (derived into strains) to balanced stresses. This weak form of balance (internal and at boundaries) allows for the autonomous identification of the constitutive law. In addition to EUCLID, other frameworks are also based on this internal and boundary force loss~ such as the DDI \revision{literature}~\cite{stainier2019model} and Equilibrium-based Convolution Neural Network (ECNN) methods~\cite{li2022equilibrium}. A modified constitutive relation error framework for learning PANN, more robust to noise impact, has also been recently developed~\cite{benady2024unsupervised,benadynnmcre}. In the ECNN approach, the model consists of strain-to-stress mapping layers (with 1$\times$1 kernels and Leaky ReLU activation functions, equivalent to a multi-layer perceptron). This constitutive model does not contain physical knowledge.

Nowadays, the current approaches in the literature train \revision{PANNs} from synthetic data. These developments are hence restricted to numerical proof of concepts. The noise included in the studies differs from real experimental noise (\eg sometimes applied on the deformation or displacement fields) and other sensor uncertainty (\eg force sensor axis uncertainty). The quantity of data used in synthetic approaches can be tuned to train the model successfully. Finally, procedures are rarely blinded; by knowing the targeted model (knowing the shape of the free energy surface, for example), the neural network architecture can hence be adequately designed using the required deformation invariants.
The closest study to the current development in the literature is~\cite{li2022equilibrium}, where the authors validate their AI model from DIC measurements in a compression test on rubber cubes. However, besides being a different constitutive model (without physical knowledge), the experiment is limited to small deformations (under $5\%$). Multiple questions are still open for the experimental training of highly constrained PANN models, architecture selection, and model validation.
\revision{Therefore, t}he identification of PANN from actual experimental data remains a challenge and has not been studied in the literature beyond 1D applications~\cite{jordan2020neural,diamantopoulou2021stress,pierre2023discovering,flaschel2023automated}. 

This article aims to train, validate, and test a PANN model within an NN-EUCLID framework based on experiments performed on hyperelastic behavior. The trained constitutive model is first optimized. 10 different architectures are evaluated and compared in terms of equilibrium loss. The best model (the reference one) is compared with a traditional Neo-Hookean model identified using the same framework. The first part of this article presents the methodology of the training pipeline, including a PANN model, the DIC measurement procedure, and the defined metrics. The second part is dedicated to the experimental setup and deformation/force data collection. Finally, the different approaches are compared with equilibrium gap metrics.

\section{METHOD}
This section presents the methodology of the EUCLID procedure (section~\ref{sec:EUCLID}) with two types of models: a traditional hyperelastic law and a physics-augmented neural network. Then, as the kinematics measurement constitutes a central piece of information, the DIC technique is described (section~\ref{sec:DIC}). Finally, comparison metrics are proposed (section~\ref{sec:METRICS}).

\subsection{EUCLID framework}
\label{sec:EUCLID}
Consider displacement field data $\bm u (\bm x)$ defined on spatial positions $\bm x$. This displacement field may come from DIC/DVC measurement procedures and is supposed to be known on the sample surface. The displacement field is expressed using a finite element mesh, composed of $N_n$ nodes, with finite element shape functions $\Phi(\bm x)$ such that:
\begin{equation}
    \bm u(\bm x) = \sum_{i=1}^{N_n} \bm u_e^i  \Phi^i (\bm x),
\end{equation}
with $u_e^i$ being the nodal displacement amplitudes. The deformation gradient field is then approximated as follows: 
\begin{equation}
     \bm F(\bm x) = \bm I + \sum_{i=1}^{N_n} \bm u_e^i  \bm \nabla \Phi^i (\bm x),
\end{equation}
where $\bm I$ is the identity matrix and $\bm \nabla$ the gradient operator defined in Lagrangian specifications.

Considering a constitutive model $\mathcal{M}$ mapping deformation $\bm F$ onto first Piola-Kirchhoff stress $\bm P(\bm F)=\mathcal{M}\left[ \bm F \right]$. 

Let a sample be loaded with $N_R$ reaction forces $R_i$ with $i\in [1,N_R]$, obtained with external force sensors.
The EUCLID framework consists of minimizing a loss function based on the force balance residuals given by a first term $\mathcal{L}_{\text{int}}$ at all free nodes and a second term $\mathcal{L}_{\text{BC}}$ at the controlled boundary conditions. The $N_n$ nodes of the finite element mesh can hence be divided into an inner part $N_{\text{int}}$ and the controlled boundary condition part $N_{\text{BC}}$, with $N_n=N_\text{int}+N_{\text{BC}}$.

In case of negligible body forces, the nodal residual can be written by integrating the stresses onto the reference domain $\Omega$:
 
\begin{equation}
    f_i = \int_{\Omega}P_{ij}\nabla_j \Phi dV = \int_{\Omega}\mathcal{M}[\bm F]_{ij}\nabla_j \Phi dV.
\end{equation}

The EUCLID framework proposes a first loss term minimizing the L2 norm on the nodal force residuals (ensuring a balanced state). 
\begin{equation}
    \mathcal{L}_{\text{int}}(t)=\sum_{i=1}^{N_{\text{int}}}\left[ f_i(t) \right]^2,
\end{equation}
However, acquisition noise assumed white and Gaussian impacts the displacement measurement (thus significantly the deformation) and may induce a bias when using the square of each nodal force residual. \revision{To reduce the noise effect, it is here proposed to modify the loss term considering the convolution of the internal nodal forces by $N_\sigma$ Gaussian kernels $G(\sigma_j)$ with a characteristic length of $\sigma_j$. The sum over multiple Gaussian kernel lengths allows not to be only focused on high frequencies that are highly impacted by the noise. 
\begin{equation}
    \hat{\mathcal{L}}_{\text{int}}(t)=\dfrac{1}{N_\sigma}\sum_{j=1}^{N_\sigma} \sum_{i=1}^{N_{\text{int}}}\left[ G(\sigma_j) \ast f_i(t) \right]^2.
\end{equation}}
With large Gaussian kernels, the equilibrium is evaluated at larger scales. 
In practice, 3 preset kernel sizes are used in the following application \revision{corresponding to 1, 2, and 3 element size}. It should be noted that pre-processing the input DIC fields to mitigate the effects of noise is a possible strategy~\cite{flaschel2021unsupervised}. However, this approach was not adopted in the current study.
The second term expects to balance the reaction forces
\begin{equation}
    \mathcal{L}_{\text{BC}}(t)=\sum_{j=1}^{N_R}  \left[ R_j(t) - \sum_{i=1}^{N_{\text{BC}}}  f_i(t) \right]^2.
\end{equation}

\revision{The complete loss consists of a weighted sum of the two terms described above, with $\gamma$ a weighting factor~\cite{flaschel2021unsupervised}. The loss also sums all $N_t$ temporal contributions such that:
\begin{equation}
    \mathcal{L}_l=\dfrac{1}{N_t}\sum_t^{N_t} \left[  \hat{\mathcal{L}}_{\text{int}}(t) + \gamma \mathcal{L}_{\text{BC}}(t)    \right].
\end{equation}}
The EUCLID method has been extended with the learning of neural-network constitutive models and is referred to as NN-EUCLID~\cite{thakolkaran2022nn}. \revision{In the latter, a weight factor was chosen: $\gamma=10$ to increase the boundary condition loss contribution.}

The EUCLID framework with the described loss will be used to train different models: 
\begin{itemize}
    \item a standard Neo Hookean material model (section~\ref{sec:NH}).
    \item multiple PANN models able to satisfy physical constraints with various neural network architectures (section~\ref{sec:PANN}).
\end{itemize}

\subsection{Constitutive models}

\subsubsection{Traditional hyperelastic model}
\label{sec:NH}
The standard model used as a reference is isotropic hyperelastic behavior following Neo-Hookean materials (NH). This model is written $\mathcal{M}_{\text{NH}}$ in the following.
\revision{In this model variant, the strain energy density function is written with the deformation invariants defined in plane stress, with
$I_1 = \text{tr}(\bm C)$ and $J = \text{det}(\bm F)$, with  $\bm C = \bm F^T \cdot \bm F$:
\begin{equation}
    W_{\text{NH}} = \dfrac{\mu}{2} \cdot (I_1 - 3) - \mu \cdot \log(J) + \dfrac{\lambda}{2}  \cdot \log(J)^2,
\end{equation}}
with $\mu$ and $\lambda$ the Lamé coefficients. 

\subsubsection{PANN architecture model}
\label{sec:PANN}
The neural network model used in this application is a PANN, $\mathcal{M}_{\text{PANN}}$,  designed for isotropic hyperelastic applications. Its architecture is based on a neural network composed of dense layers that take as inputs the deformation tensor $\bm F$ and outputs the first Piola-Kirchhoff stresses $\bm P$. Various physical constraints are implemented following reference papers in the literature~\cite{klein2022polyconvex,as2022mechanics,linden2023neural}:
\begin{itemize}
    \item Manual definition of the deformation invariants before the first layer. The deformation invariants are defined as
$I_1 = \text{tr}(\bm C)$, $I_2 = \frac{1}{2}\left(\text{tr}(\bm C)^2-\text{tr}(\bm C^2)   \right)$ and $J = \text{det}(\bm F)$, with  $\bm C = \bm F^T \cdot \bm F$ are used as input of the first layer. 
    \item Thermodynamic consistency: the first Piola-Kirchhoff stress is obtained by derivation of the strain energy density function $W$: $\bm P(\bm F)=\dfrac{\partial W(\bm F)}{\partial \bm F}$. An intermediate output of the Neural Networks is hence a scalar value $W$.
    \item Stability and Positivity of the potential is obtained by the polyconvexity with respect to $\bm F$ and det$(\bm F)$ by using the convex \emph{softplus} activation functions for the hidden layers: $\mathcal{A}(y)=\log(1+e^y)$ \revision{and positive model weights $\bm \omega>0$. Note that, for the initial selection of deformation invariants, both $J$ and $-J$ are implemented to avoid clipping to only positive values. The relationship between two layers can hence be written, with $\bm \omega^i$ and $\bm b^i$ the weights and bias of layer $i$, and $\bm z^{i-1}$ and $\bm z^{i}$ the neuron values of layers $i-1$ and $i$:
    \begin{equation}
        \bm z^{i}=\mathcal{A}(\bm \omega^i\bm z^{i-1}+\bm b^i)
    \end{equation}
    \item Normalization of the energy and stresses: $\bm P(\bm F=\bm 1)=\bm 0$, and $W(\bm F=\bm 1)= 0$. This normalization is obtained by performing the forward pass of a non-deformed state $\bm F=\bm 1$ into the model and subtracting it from the output energy and stresses. This computation is performed at each batch.}
\end{itemize}

\begin{figure}[t]
\centering
     {\includegraphics[width=0.8\textwidth]{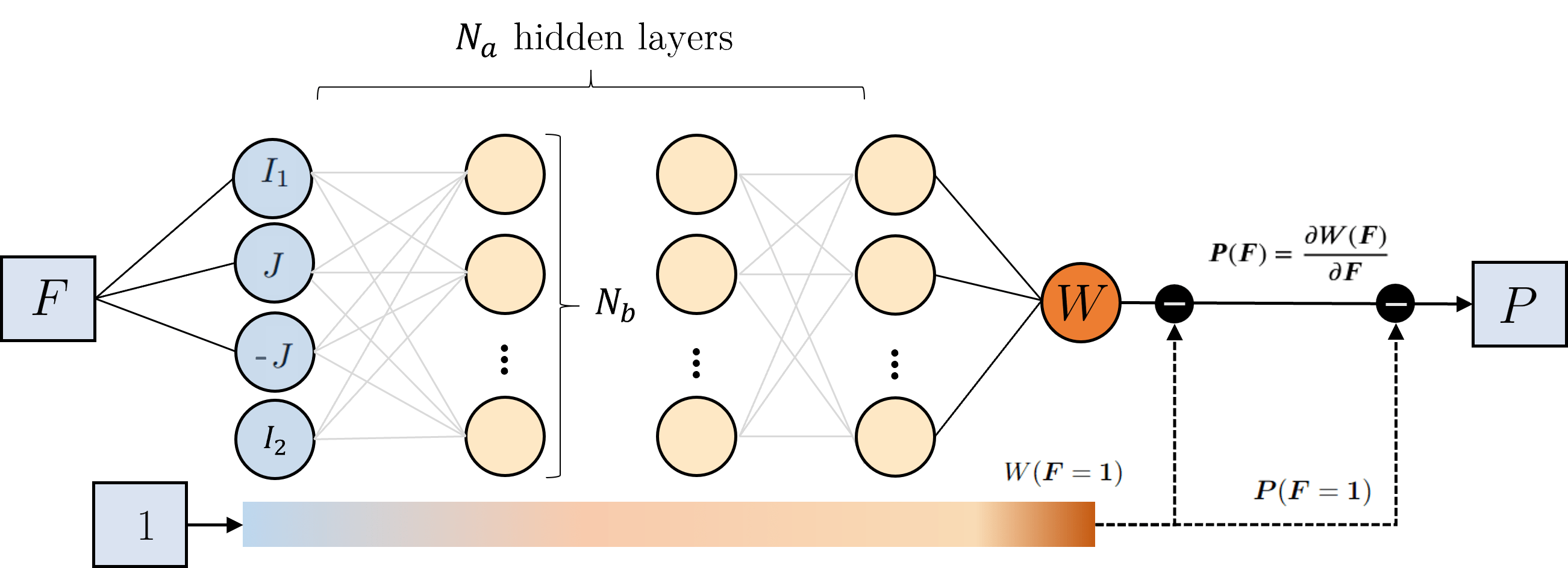}}
	\caption{PANN architecture composed of: $N_a$ hidden layers composed of $N_b$ neurons}
	\label{fig:PANN} 
\end{figure}

A summary of the model architecture is presented in figure~\ref{fig:PANN}.
The core architecture comprises a first hard-coded layer to compute the deformation invariants. This "pre-processing" is followed by $N_a$ dense hidden layers, each composed of $N_b$ neurons and \emph{softplus} activation functions. The last top layer links the $N_b$ neurons to the scalar energy value with a linear activation function. 
Without knowing the true material behavior, the model architecture has to be optimized. Model architecture optimization is a vast subject in neural networks.
This application evaluates different PANN architectures from the loss obtained on the validation set. Table~\ref{tab:arch} and associated with models named PANN\_$N_a(N_b)$ shows the number of tested hidden layers and neurons. Multiple architectures are tested, with $N_a \in  \{2, 5, 10\}$ hidden layers and  $N_b \in  \{8, 16, 32, 64\}$ neurons per layer. This represents an extensive range of model sizes, from 145 trainable parameters (weights and biases) up to 17k.
In comparison, it is important to recall that the chosen NH model comprises only two parameters.
\begin{table}[t]
\centering
\begin{tabular}{lccr}
\toprule
Model name   & \# layers ($N_a$) &\# neurons ($N_b$) & \# parameters \\ \midrule
PANN\_2(8)   & 2                        & 8                         &       145               \\
PANN\_2(16)  & 2                        & 16                        &       361               \\
PANN\_2(32)  & 2                        & 32                        &       721               \\
PANN\_5(8)   & 5                        & 8                         &       417               \\
PANN\_5(16)  & 5                        & 16                        &       1,233             \\
PANN\_5(32)  & 5                        & 32                        &       2,593             \\
PANN\_10(8)  & 10                       & 8                         &       1,345             \\
PANN\_10(16) & 10                       & 16                        &       4,513             \\
PANN\_10(32) & 10                       & 32                        &       9,793             \\ \midrule
PANN\_5(64)  & 5                        & 64                        &       17,217            \\
\bottomrule
\end{tabular}
\caption{Different tested PANN architectures}
\label{tab:arch}
\end{table}

The optimization is performed with an ADAM optimizer~\cite{kingma2014adam} with a learning rate following a predefined exponential decay schedule at each epoch $k$: $l_r(k)=10^{-4}\cdot 0.95^{(-\frac{k}{112})}$. This corresponds to a start at $10^{-4}$ and a decrease of around 1/10 every 10k epoch. The learning rate schedule was calibrated from the model with the largest number of parameters and used for all other models. A significant gain in the convergence time could be achieved by optimizing this schedule for each architecture.

Weights are initialized with random positive values (uniform distribution in [0-0.1]). For the training process, each epoch is composed of $N_t$ batches, each consisting of a single loading step for a particular geometry. The loading steps are shuffled at the start of every epoch to ensure variability.
The neural network architecture is developed using the TensorFlow library in Python.

\subsection{Digital image correlation}
\label{sec:DIC}
Both previous approaches require to have access to kinematics measurements. With plane stress assumption, 2D full-field measurements can be carried out on the sample surface.
A global DIC approach~\cite{hild2012comparison} is used to measure the 2D displacement field. It consists of the registration of an image $f(\bm x)$ in the reference configuration and a series of pictures $g ( \bm x, t )$ in the deformed configuration indexed by time $t$. The DIC algorithm minimizes the sum of squared differences between the deformed image, corrected by the measured displacement field at time $t$: $\bm u ( \bm x, t )$, and the reference image
\begin{equation}
    \Gamma_f^2(t) = \dfrac{1}{2\sigma_f^2\|\Omega\|}\sum_{\bm x \in \Omega} \left(    g(\bm x+\bm u(\bm x,t),t)-f(\bm x)     \right)^2.
\end{equation}
Global DIC consists of describing the search kinematics by finite element shape functions $\Phi_i(\bm x)$ and degrees of freedom $\bm u_e$ such that $\bm u(\bm x,t) = \sum_{i=1}^{N_n} \bm u_e^i(t)  \Phi^i (\bm x)$, where $\bm x \in \Omega$ are the considered pixel coordinates in the region of interest and $\sigma_f$ the standard deviation (expressed in grey levels) of the Gaussian white noise assumed to affect each image independently (including the reference one, which is responsible for the factor 2). \revision{The mesh chosen for the DIC measurement (and the shape functions) $\bm \Phi$ are the same as the one used in the PANN procedure. This is one of the main advantages of global DIC, as it avoids applying interpolation to match different kinematics supports. DIC measurements and PANN computation are hence completely merged.}
In addition to the displacement field measurement, brightness and contrast correction are also applied to compensate for the non-conservation of the optical flow~\cite{sciuti2021benefits}. 

The minimization of this functional, with respect to the displacement, brightness, and contrast degrees of freedom, is solved by successive linearizations and corrections using a modified Gauss-Newton scheme. A multi-scale approach is performed, both in the kinematical model (with mechanical regularization~\cite{mendoza2019complete}) and image downsizing~\cite{buljac2018digital}. The optimization starts with coarse scales and finishes with raw full-resolution images without mechanical regularization. 
DIC is performed based on the DIC software \emph{Correli}~\cite{leclerc20153}.

\revision{As DIC is a 2D measurement procedure, a plane stress assumption has to be made concerning the depth behavior, with negligible stress in the depth direction. }

\subsection{Evaluation metrics}
\label{sec:METRICS}
Specific error values can be designed to evaluate the quality of the identified model. 

The first monitored error is the previously defined training loss $\mathcal{L}_{\text{l}}(t)$ on the training set, validation, and test datasets. Based on the validation dataset, this loss will be used to select the best PANN model from the different tested architectures.

The total loss can be split into two terms, called metrics in the latter, to simplify the distinction with the training loss: $\mathcal{L}_{\text{int}}(t)$, and $ \mathcal{L}_{\text{BC}}(t)$. The first term represents the internal equilibrium, and the second part represents the boundary condition error. Evaluating those two metrics individually will give more information on the model performances.

\section{DATA}

\subsection{Experiments}

Two uni-axial tensile tests have been performed on a \revision{neoprene hyperelastic rubber material with unknown a-priori properties as stored in uncontrolled environment conditions}. \revision{The loading sequences were performed on two different tested geometries: a sample with a hole and a second full sample without a hole. In the ``With Hole'' test}, the shape of the sample was cut with a hole to enhance a heterogeneous loading and increase the mechanical content of the measurement.

\begin{figure}[t]
\centering
        \subfigure[Geometries and DIC meshes for \revision{experiment - ``With Hole''} (left) and \revision{experiment - ``No Hole''} (right)]{\includegraphics[width=0.4\textwidth]{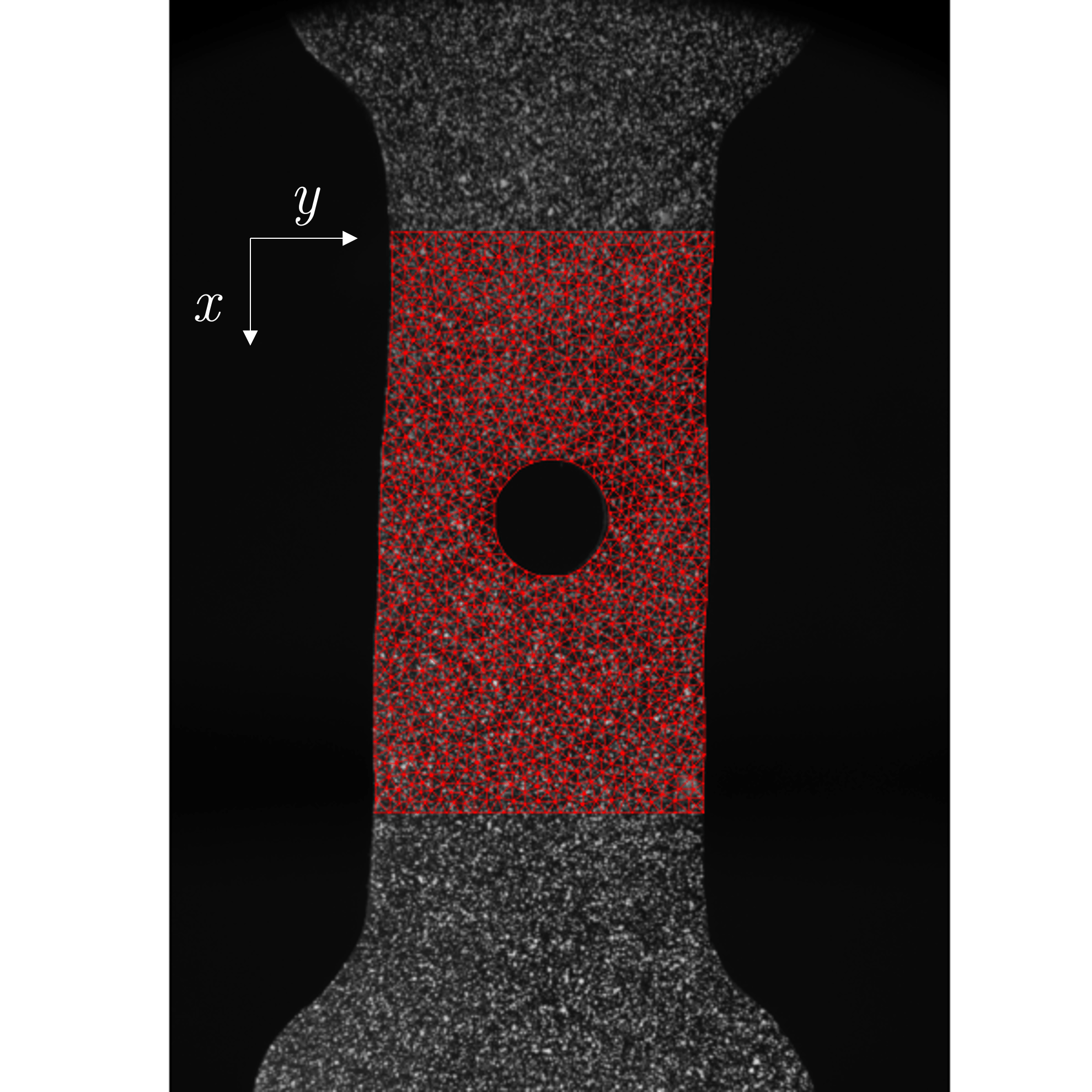}  \includegraphics[width=0.4\textwidth]{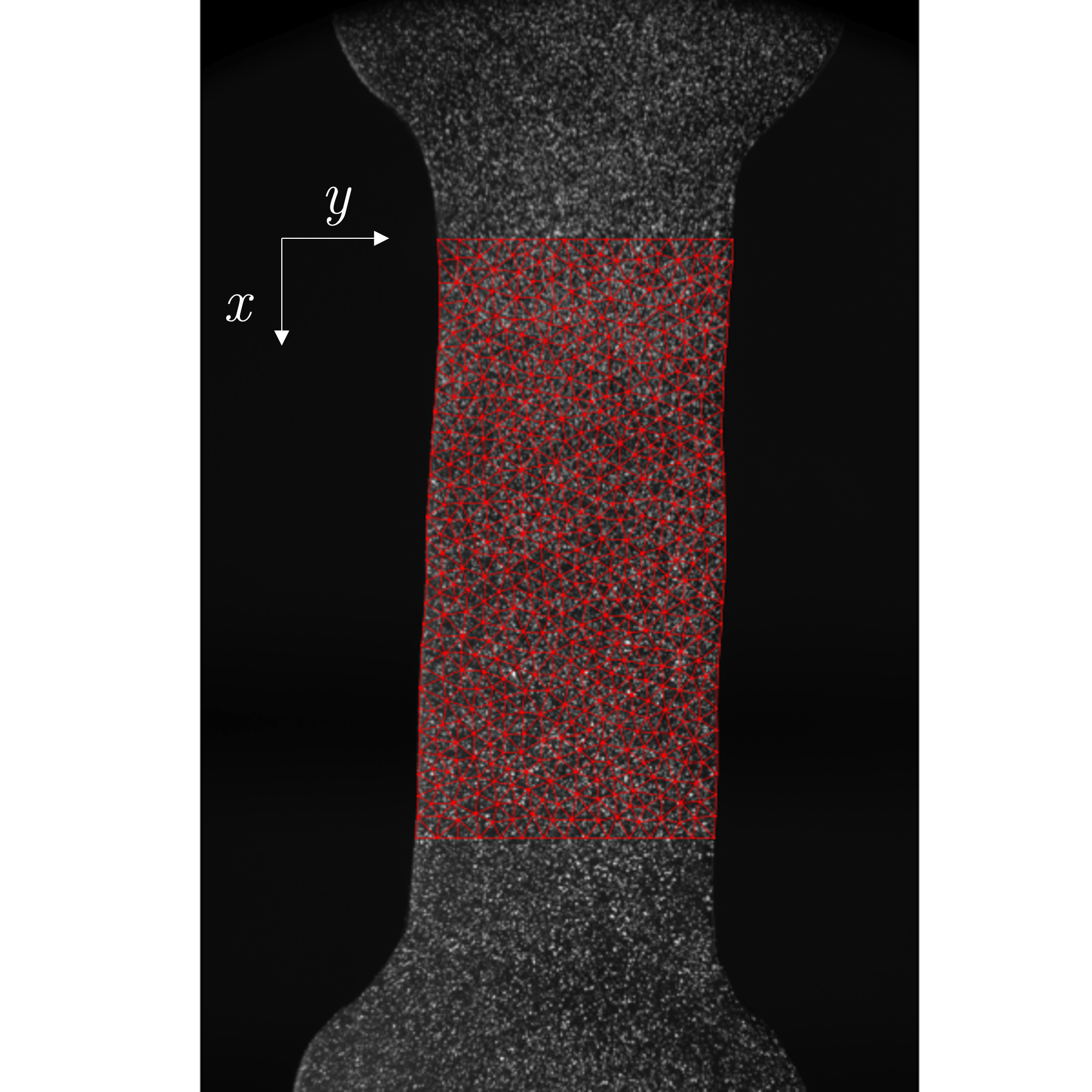} }
        
        \subfigure[Force stroke curves  for \revision{experiment - ``With Hole''} (left) and \revision{experiment - ``No Hole''} (right)]{\includegraphics[width=0.4\textwidth]{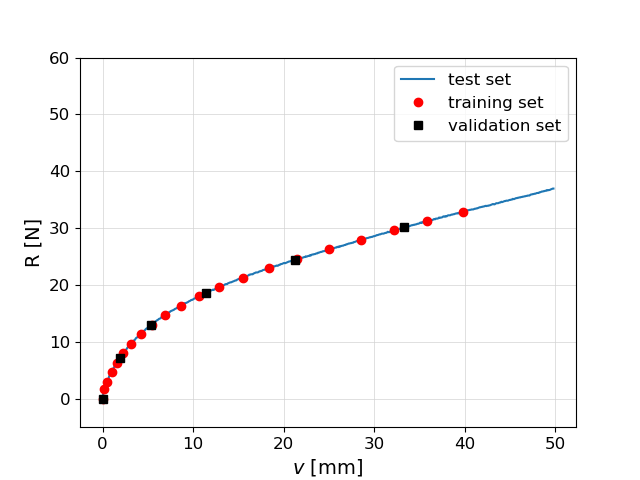} \includegraphics[width=0.4\textwidth]{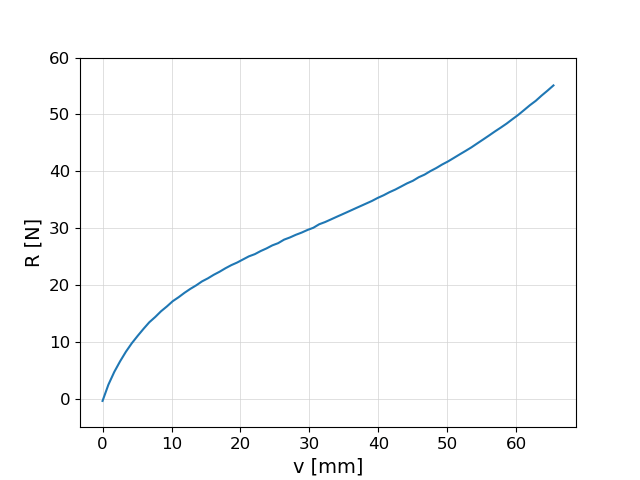}  }

	\caption{Samples, meshes, and force/displacement curves measured by the loading machine for \revision{experiment - ``With Hole''} (a) and \revision{experiment - ``No Hole''} (b). The red circles in (b) indicate the 20 training data and the black squares represent the validation data. All other blue steps are used for the test. 
     }
	\label{fig:loading} 
\end{figure}

The imaged surface was covered with thin white paint speckles to create an image gradient used for the DIC measurement. As the sample color was initially black, no black paint was used (allowing for a reduction in the projected paint quantity and thus reducing the risk of paint cracking at large deformation). The shapes and appearances of the samples are shown in figure~\ref{fig:loading}(a).

The loaded surface was imaged using a PCO-edge camera with a telecentric lens (diameter 125~mm) and LED panels. At this scale, each pixel has a resolution of 78~$\mu$m. \revision{The electronic noise estimated standard deviation is: $\sigma_{f1}=374$~gray levels in the experiment - ``No Hole'', and $\sigma_{f2}=340$~gray levels in the experiment - ``With Hole'' for 16-bit GL images and an image intensity covering most of the 65,536 possible gray level values}).
The uni-axial testing machine was a thermo-regulated electro-mechanical INSTRON system equipped with a 5~kN loading cell (estimated standard deviation $\sigma_R=0.020$~N). The sample was continuously loaded during the test with a displacement control. The stroke of the machine is written $v$.
The material was initially cycled \revision{ten times, at maximal load to reach a stabilized behavior. The stabilized cycles appeared already at cycle 3.}

\paragraph{\revision{Experiment - ``With Hole''} - The geometry contains a centered 10 mm diameter hole.} 30 images were taken in the reference unloaded state for DIC uncertainty quantification. Then, the sample was continuously loaded, with a controlled displacement of 5 mm/min until reaching 50 mm, and an image acquisition every second, allowing for fine temporal sampling. In total, 551+30 images were acquired during the experiment. 

\paragraph{\revision{Experiment - ``No Hole''} - } The first geometry corresponds to a standard uni-axial sample (without hole) with an effective surface of 55$\times$26~mm$^2$ and a thickness of 2~mm. The sample was loaded with a displacement control of 5~mm/min until a maximal value of 67.5~mm and an image acquisition every 10 seconds. During \revision{experiment - ``No Hole''}, a total of 81 images were acquired. \revision{It is important to note that in practice, both tests were conducted on the same rubber sample. The first test, "No Hole," was performed first chronologically. The ``With Hole'' test was conducted later, using the same experimental setup but on a different day with different paint speckles. Consequently, the constitutive behavior is very similar for the two experiments.}

\begin{figure}[t]
\centering
     {\includegraphics[width=1.\textwidth]{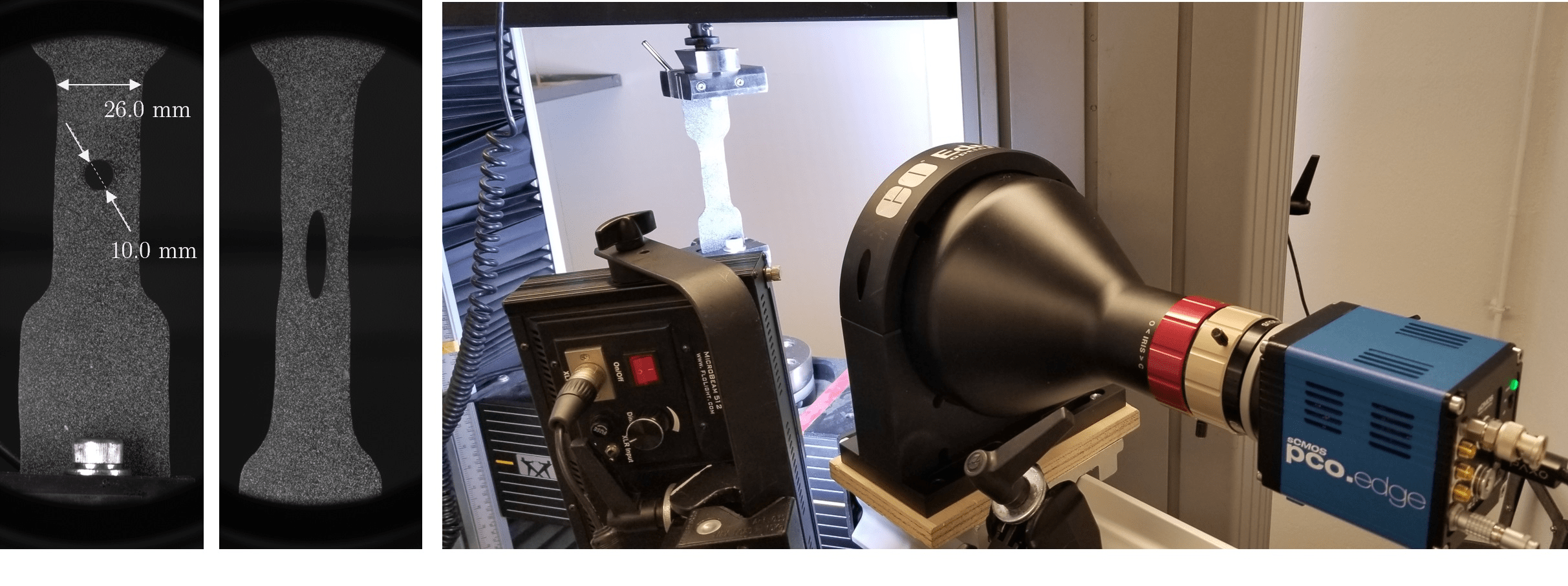}}
	\caption{Experimental setup with the images at reference (step 1, $R(0)=0$~N) and at the end of the loading (step 581, $R(581)=36.70$~N) for the \revision{experiment - ``With Hole''}.}
	\label{fig:exp_setup} 
\end{figure}

\begin{figure}[t]
\centering
     {\includegraphics[width=1.\textwidth]{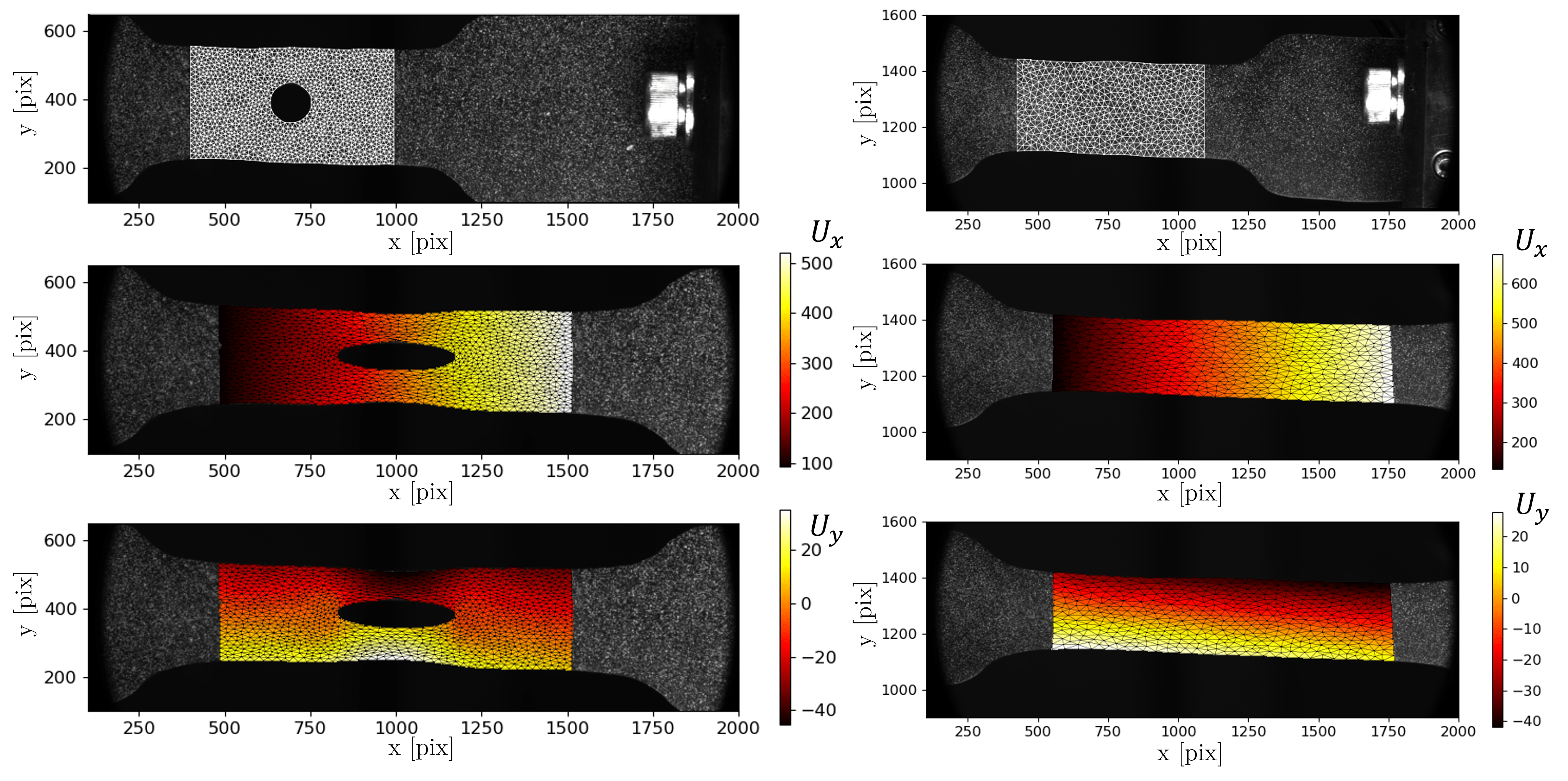}}
	\caption{Reference and final images for \revision{experiment - ``With Hole''} (left) and \revision{experiment - ``No Hole''} (right) with the measured displacement field, expressed in pixels.}
	\label{fig:exp_setup_DIC} 
\end{figure}

The first and final acquired images for \revision{experiment - ``With Hole''} are presented in figure~\ref{fig:exp_setup} with the loading and acquisition global setup. The important sample deformation can be seen in the second picture. Figure~\ref{fig:exp_setup_DIC} shows the measured displacement field represented in the deformed configuration of the final loading steps for the two experiments.

\subsection{Data split}
The dataset was split into different subsets to train, validate, and test the model. 
The training set is used to learn the model weights. The validation set is to monitor the model evolution, evaluate the under-/over-fitting, calibrate the hyperparameters, and potentially select the best model (early stopping). Finally, the test data are used for the evaluation to generalize on unseen loading conditions. 

\revision{Experiment - ``With Hole''} was divided into training, validation, and test parts, while \revision{experiment - ``No Hole''} was dedicated to the test. 
Table~\ref{tab:split} presents a detail of the data split. 

\paragraph{Training:} 20 loading steps in \revision{experiment - ``With Hole''}, with a regular force sampling in [0 -- 31.5]~N were chosen for the training part, specifically steps [32, 35, 41, 47, 55, 64, 76, 89, 105, 124, 146, 171, 199, 231, 265, 303, 342, 381, 421, 465]. For an additional evaluation, the model will also be trained with [3,6,10,20,40] training data, chosen with the same linear sampling method.
\paragraph{Validation:} 6 loading steps in \revision{experiment - ``With Hole''}, with a regular force sampling, were chosen for the validation part. Selected steps are at steps [1, 51, 88, 155, 262, 394] and correspond respectively to an applied force of [$0, 6.75, 12.5, 18.25, 24.0, 29.75$]~N. This dataset will be used to monitor the model training evolution, compare the different hyperparameters ($N_a$ and $N_b$), and select the best reference model.
\paragraph{Test:} The remaining 554 steps of \revision{experiment - ``With Hole''} were used for the validation to verify the capacity of the model to generalize. It can be noted that the last steps were only selected for the validation to evaluate the model's ability to extrapolate with a higher material state.
The 80 data steps of \revision{experiment - ``No Hole''} were exclusively used for the test. Having a distinct sample exclusively for the test allows the evaluation of the model to generalize on unseen data: different mechanical states, mesh, sample geometry, and paint speckles.

\revision{
\begin{table}[t]
\centering
\begin{tabular}{lcc}
\toprule
\revision{Experiment}        & \revision{``With Hole''} &  \revision{``No Hole''}  \\ \midrule
Maximal stroke                & 50 mm                & 70 mm                     \\
Number of images             & 551 (+30 references) & 81                        \\
Training steps               & 20                   & -                          \\
Validation steps             & 6                    & -                           \\
Test steps                   & 554                  & 80                             \\ \midrule
Nodes number                 & 1288                 & 567                           \\
Element number (T3)          & 2422                 & 1048                      \\\bottomrule
\end{tabular}
\caption{Summary of the two experiments' characteristics and data split}
\label{tab:split}
\end{table}}

The force/displacement curves (as measured by the machine) are presented in figure~\ref{fig:loading}(b) with a highlight on the different training (red dots), validation (black squares), and test steps (blue).

\section{RESULTS}

\subsection{DIC results}
The DIC mesh was positioned pixel-wise from the images \revision{by applying image gray level threshold techniques} to follow the sample geometry precisely. \revision{This segmentation was performed using a simple gray level threshold as the sample geometry can clearly be identified from the black background.} An accurate mesh positioning is required to evaluate the static problem correctly. \revision{After the pîxel-wise thresholding of the sample surface, the mesh was generated with GMSH software}~\cite{geuzaine2009} and an average element size of 25~mm for the geometry of \revision{experiment - ``No Hole''} and 15~pixels for \revision{experiment - ``With Hole''}, corresponding respectively to 567 nodes with 1048 T3 elements, and 1288 nodes with 2422 T3 elements. The ``With Hole'' geometry was meshed with finer elements to better identify the heterogeneous deformation state around the hole.

\begin{figure}[t]
	\centering
        \subfigure[Displacement fields]{\includegraphics[width=0.3\textwidth]{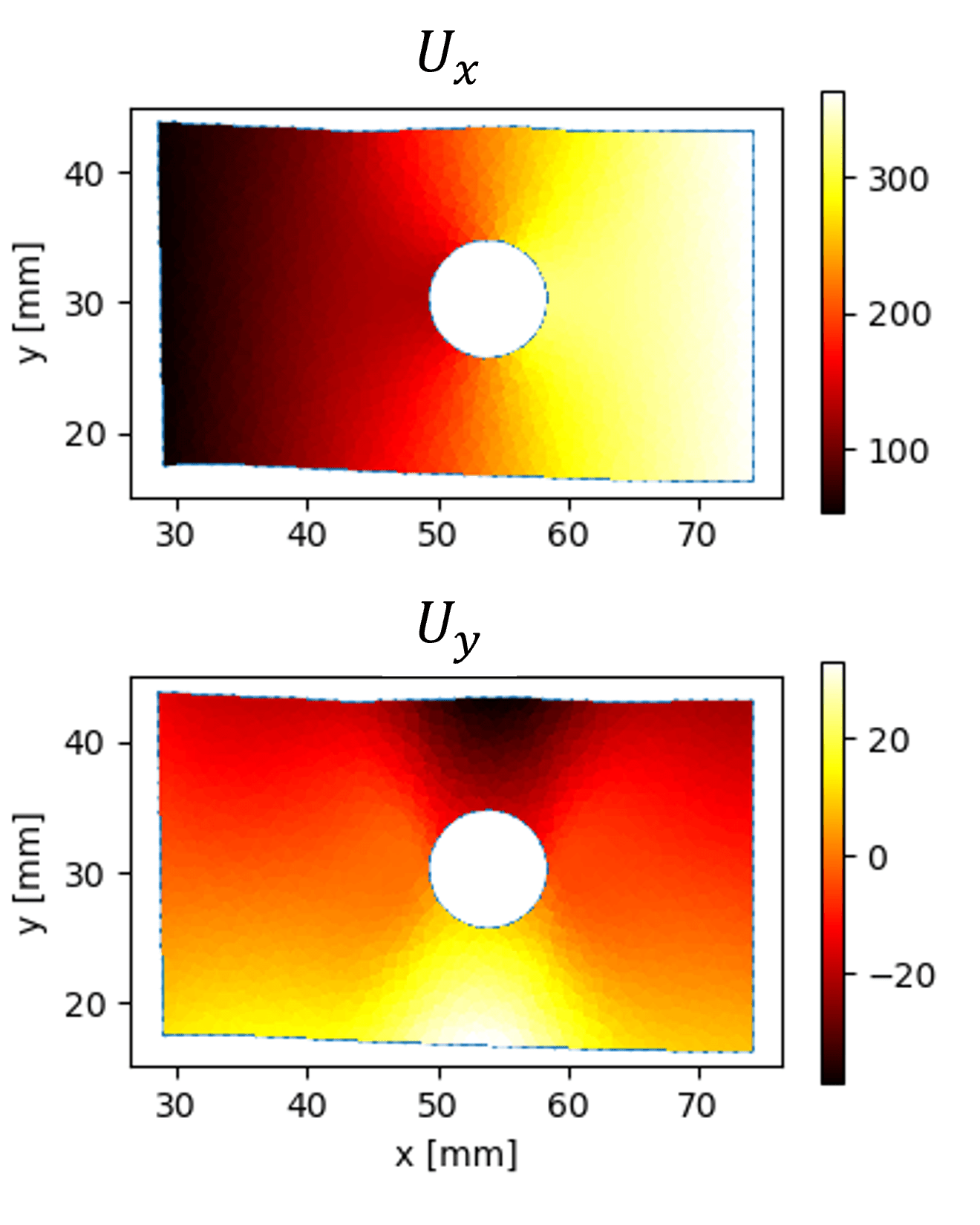}}
        \subfigure[Deformation tensor]{\includegraphics[width=0.54\textwidth]{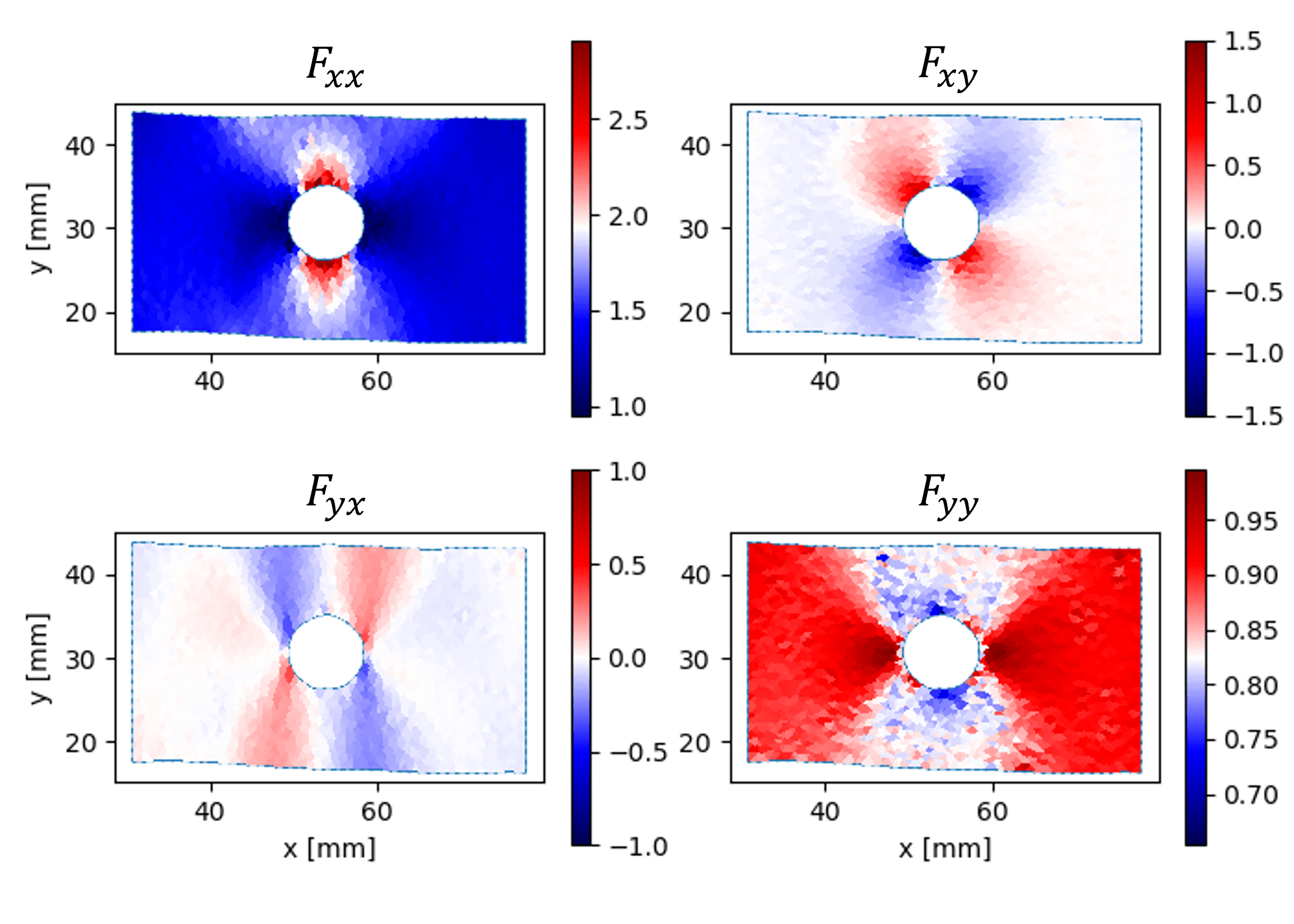}}
	\caption{DIC results for step 421 ($R(421)=31.2$~N): (a) displacement field, expressed in pixels and (b) deformation tensor.}
	\label{fig:421} 
\end{figure}

DIC was applied using the first image as the reference and all subsequent deformed ones. It converged on all steps, leading to a final root mean square residual value starting at 1\% of the image gray level dynamic for the first steps (where the loading was low) and increasing progressively to 2.6\%. This residual field increase may be due to paint issues that may not perfectly follow the material at the important deformation states. All measurements allow for storing a $\{R,\bm F\}$ database. The displacement field on the nodes of the mesh and deformation tensor, expressed for each element of step 421, is shown in figure~\ref{fig:421}. A significant axial displacement of 350 pixels is measured (equivalent to 25~mm) with a large gradient, leading to $F_{xx}$ values reaching more than 2.5 (corresponding to 150~\% of tension deformation) around the hole.  
At the maximum loading, $t=580$, the axial deformation exceeded 200$~\%$. It can be noted that this deformation amplitude is not classical (\eg compared to 5\% deformation in~\cite{li2022equilibrium}) and allows exhibiting a complex non-linear behavior with large displacements.

\begin{figure}[t]
	\centering
        \includegraphics[width=0.75\textwidth]{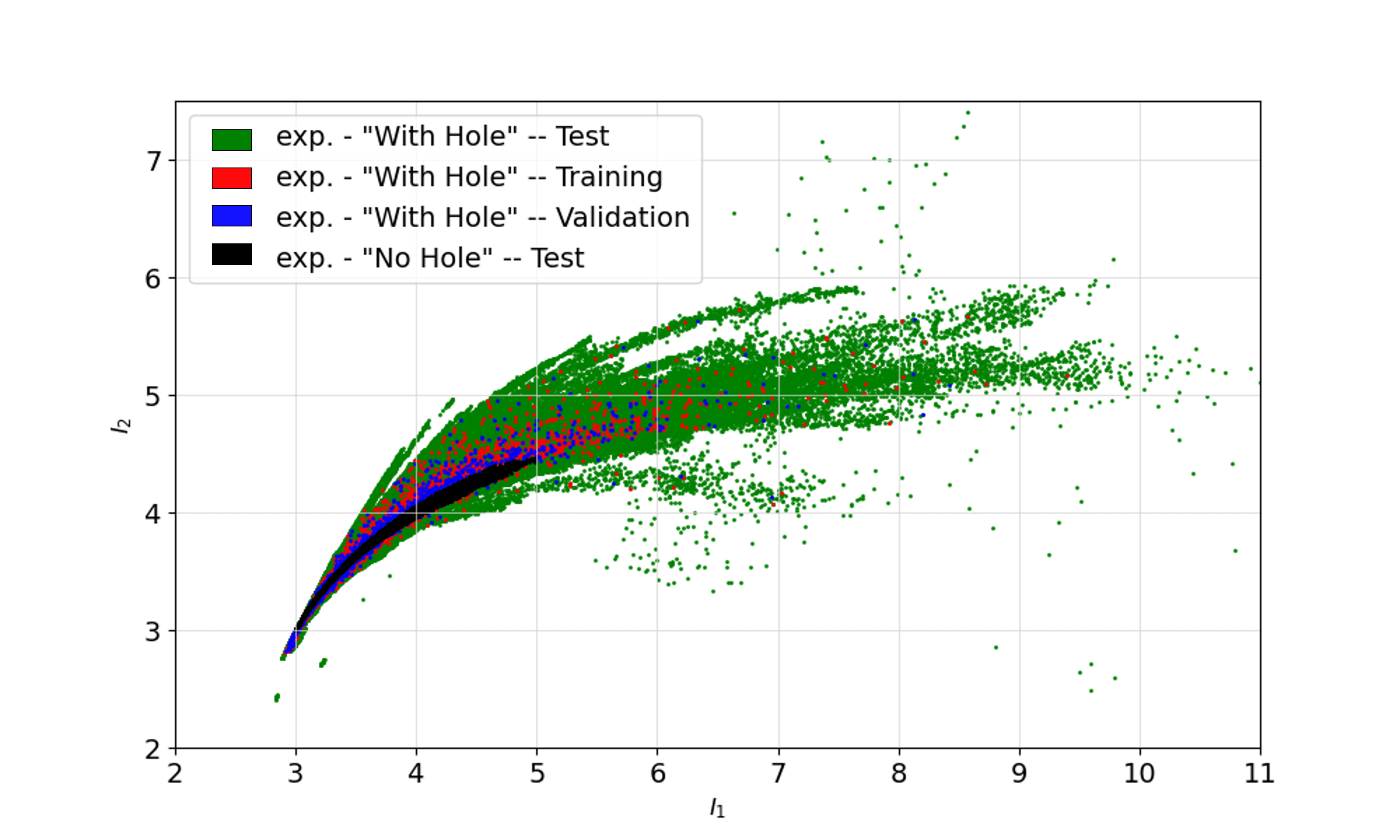}
	\caption{\revision{Deformation dataset expressed in the $(I_1,I_2)$ invariant space.} 
        }
	\label{fig:inv} 
\end{figure}

\revision{The deformation tensor can also be represented in the invariant space $(I_1,I_2)$. Figure~\ref{fig:inv} presents the $I_1$ and $I_2$ invariants of the two tests as obtained by DIC. }
\revision{The green markers represent experiment - ``With Hole'' for the testset with 1,402,338 points (number of elements in all loading steps), red markers experiment - ``With Hole'' for the training part (50,862 points), and the blue markers experiment - ``With Hole'' for the validation part (14,532 points), and the gray markers experiment - ``With Hole'' for the test (80,696 points). }
Few outliers are visible in the graph. These are due to inaccurate displacement field measurements (thus deformation) located at the hole's edge or at the boundary condition. Although it appears clearly in the graph, it corresponds to only a few elements compared to the entire 1.4 million elements in the test of \revision{experiment - ``With Hole''}. For future experiments and if the noise became too important, the filtering procedure could be included as a post-processing of the DIC field to remove those outliers (as proposed in the initial EUCLID framework~\cite{flaschel2021unsupervised}). 

\begin{figure}[t]
	\centering
        \includegraphics[width=0.75\textwidth]{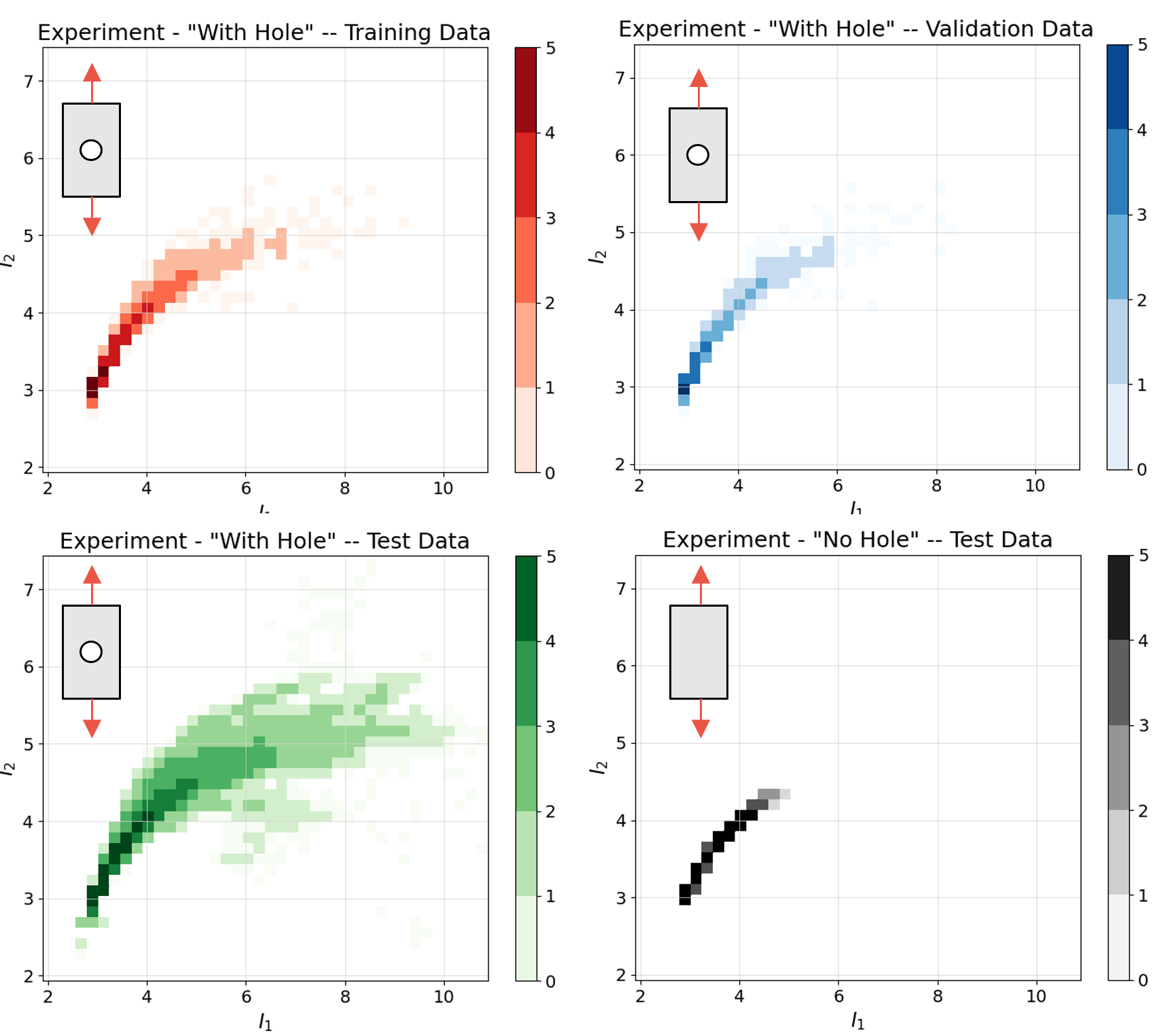}
	\caption{\revision{Distribution of the data in the invariant space from the different datasets and experiments. The color represents the occurrence, expressed in a log10 scale and computed with a common bins grid.}}
	\label{fig:inv2} 
\end{figure}

Figure~\ref{fig:inv2} represents the data distribution for the different datasets. The distribution was computed from a common 40$\times$40 \revision{equally sampled bins grid in $[I_1,I_2] \in [\{2,11\},\{2,7.5\}]$ and the occurrence of points was displayed using a log10 color bar scale.}. With much fewer element numbers, \revision{experiment - ``No Hole''} also has a much smaller covered area. Training and validation cover a similar area (although validation comprises only 6 steps and training 20). Finally, the test set of \revision{experiment - ``With Hole'' covers a large area in the $(I_1,I_2)$ space.}
Although \revision{experiment - ``With Hole''} was performed with a hole and in large displacements, the informative content is essentially represented with 1D scattered beam manifolds. The experiments are not highly multiaxial and do not exhibit complex mechanical states.

\subsection{Model selection}
\begin{figure}[t]
\centering
        \includegraphics[width=0.55\textwidth]{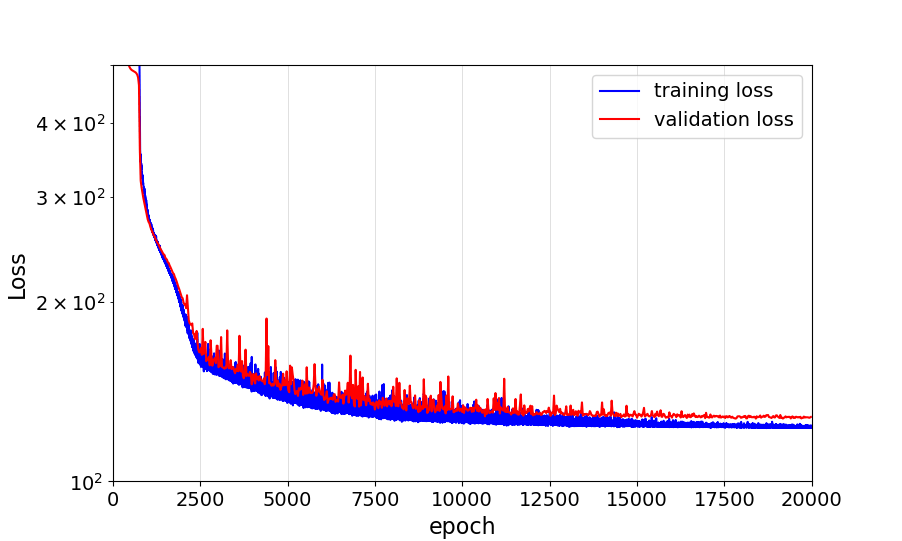}
	\caption{Convergence of the training loss (blue) and validation loss (red) for the reference PANN\_5(16) model.}
	\label{fig:conv} 
\end{figure}

The model was trained during 20k epochs with the training database of 20 loading steps of \revision{experiment - ``With Hole''} and validated every 25 epochs with the 6 loading steps of \revision{experiment - ``With Hole''}. The loss starts at values around $10^4$ and converges to 121. For comparison, the loss estimated on the 30-first unloaded DIC computation gives 4.2 (std 0.22). For the model "PANN\_5(16)", the training took approximately 10 minutes on a laptop with an 8~Gb GPU unit (Nvidia RTX-A2000 Mobile) per considered training step. Thus, it corresponds to approximately 3h30 for the 20 considered loading steps.
The convergence curve on the training and validation sets for the PANN\_5(16) model, considered the reference one, is presented in log-scale in Figure~\ref{fig:conv}. The first significant loss decrease happens in 2500 epochs (thus 50k model back-propagations). Then, the slope is reduced, and the final 2.5k-12k epochs allow for a reduction from 160 to 121. Finally, from 12k to 20k epoch, the loss remains steady. 

\begin{figure}[t]
	\centering
        \subfigure[Training loss]{\includegraphics[width=0.4\textwidth]{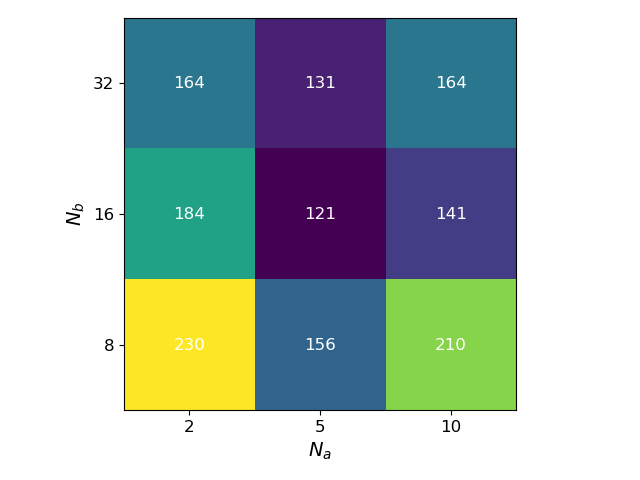}} 
        \subfigure[Validation loss]{\includegraphics[width=0.4\textwidth]{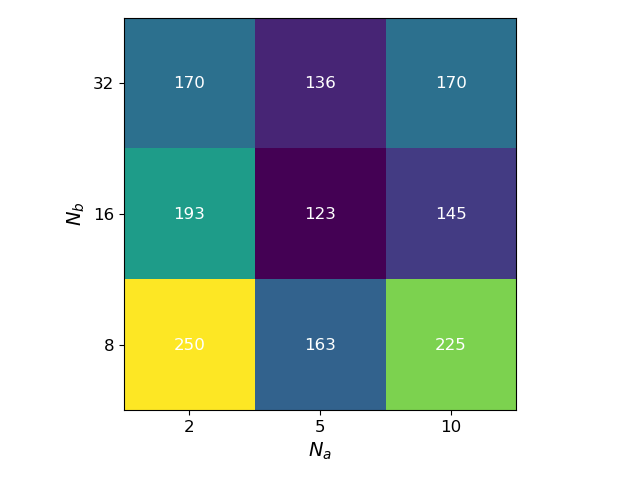}}
	\caption{Final training and validation losses for the different model architectures composed of $N_a$ hidden layers and $N_b$ neurons.}
	\label{fig:res_train} 
\end{figure}

\paragraph{Model architecture selection}
The training and validation loss results for the different model architectures are presented in Figure~\ref{fig:res_train}. Those results correspond to the loss at the last training epoch. Although it is a common practice to select the best model from the validation set (early stopping), all validation curves were essentially monotonic (without considering high-frequency variations at each epoch) with a final stabilized and minimal loss value. 
\begin{itemize}
    \item $N_a=2$: Models composed of only two hidden layers have higher equilibrium loss than the other architectures, regardless of the number of neurons. 
    \item $N_a=5$: The lowest loss values are obtained with 5 hidden layers. The best architecture (with respect to the loss criterion) is obtained with 5 hidden layers and 16 or 32 neurons per layer, model PANN\_5(16). When increasing the number of neurons to 64 (not shown in the figure), the model score does not improve (training: 146, validation 152). For this last architecture, convergence was very noisy. 
    \item $N_a=10$: The performances do not increase with more hidden layers.
\end{itemize}

In the latter, the model PANN\_5(16), composed of 5 hidden layers and 16 neurons per layer (with 1233 trainable parameters), will be considered as the reference model and will be compared with the NH model.

To evaluate the variations due to the stochastic training, the reference model was trained 5 times from scratch. The standard deviation computed on the 5 final losses is 2.8.

\begin{figure}[t]
	\centering
        \includegraphics[width=0.45\textwidth]{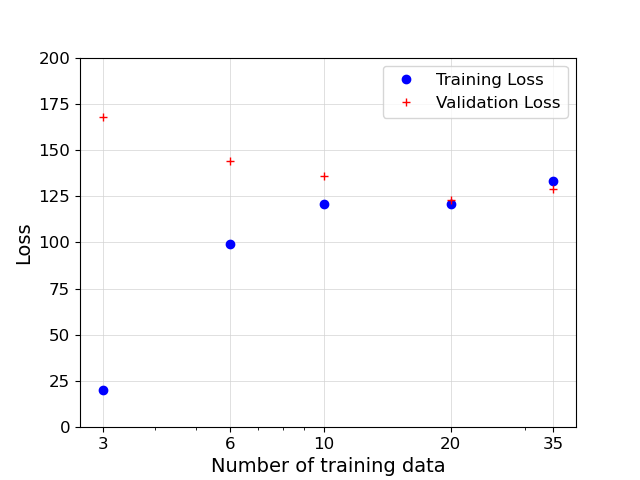}
	\caption{Converged model loss with the PANN\_5(16), computed on the training and validation dataset, from different training dataset sizes.}
	\label{fig:traindata} 
\end{figure}

\paragraph{Number of training data}
The evaluation of the model learning (PANN\_5(16)) with a different number of training data is evaluated [3, 6, 10, 20, 40]. For each evaluation, the model is trained from scratch. Results are presented in the figure~\ref{fig:traindata} for the training and validation data. Note that while the training data changes, the validation dataset is the same for all evaluations to compare results.
With few training data \revision{($<10$)}, the model performs well on the training steps yet presents a significant error on the validation data. With more training steps, the training and validation show similar results and stabilize around 125.
For the following section, 20 training data were used.

\subsection{Comparison PANN\_5(16) with NH}
The EUCLID procedure was applied to identify the two material parameters of the NH model. At convergence, the obtained parameters were $E_{\text{NH}}=1.74$~MPa and $\nu_{\text{NH}}=0.471$. The final converged value of the loss was $299$. This equilibrium loss is much higher than the previous PANN results.

For analysis purposes, the PANN model was trained with synthetic data generated from the identified NH behavior, with the geometry and boundary conditions extracted from the actual experiment. The PANN is able to reproduce the NH behavior with a converged loss of 0.1. \revision{This training was performed to verify if the chosen PANN model could at least reproduce the NH model. This training was then removed, and all later training started from scratch.}

\begin{figure}[t]
	\centering
        \subfigure[Stresses with the NH model $\mathcal{M}_{\text{NH}}(\bm F)$     ] {\includegraphics[width=0.45\textwidth]{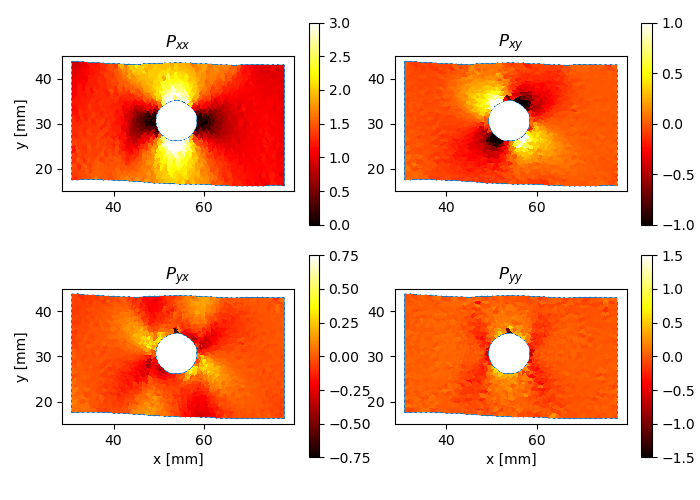}} \hspace{1cm}
        \subfigure[Stresses with the PANN model $\mathcal{M}_{\text{PANN}}(\bm F)$ ]{\includegraphics[width=0.45\textwidth]{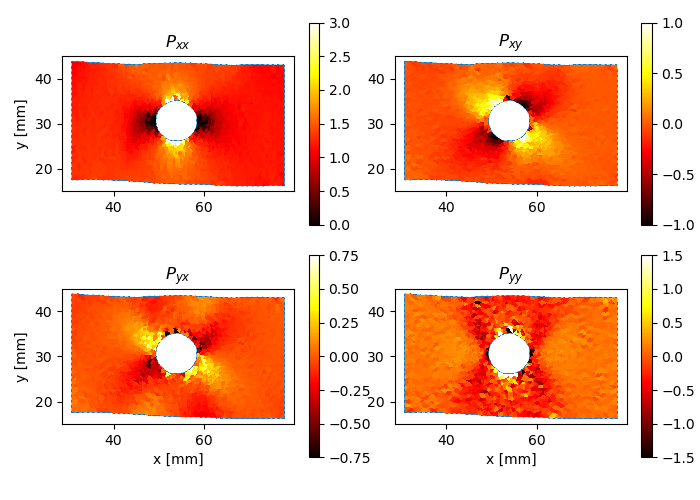}}
	\caption{Stresses predicted for step 500, expressed in MPa. 
                }
	\label{fig:P451} 
\end{figure}

The first Piola-Kirchhoff stresses extracted for the two models are presented in Figure~\ref{fig:P451}.
The two models predict the shear components $P_{yx}$ and $P_{xy}$ with similar shapes and amplitudes. Large differences can be seen in the fields $P_{xx}$ and $P_{yy}$. For the first one, the amplitudes differ with higher values for the NH model. An important predicted axial force difference is hence expected from the two models. For the field $P_{yy}$, the NH model presents incorrect behavior on the large compression area around the hole as it predicts positive stress values.

\begin{figure}[t]
	\centering
        \subfigure[\revision{Experiment - ``With Hole''}]{\includegraphics[width=0.45\textwidth]{  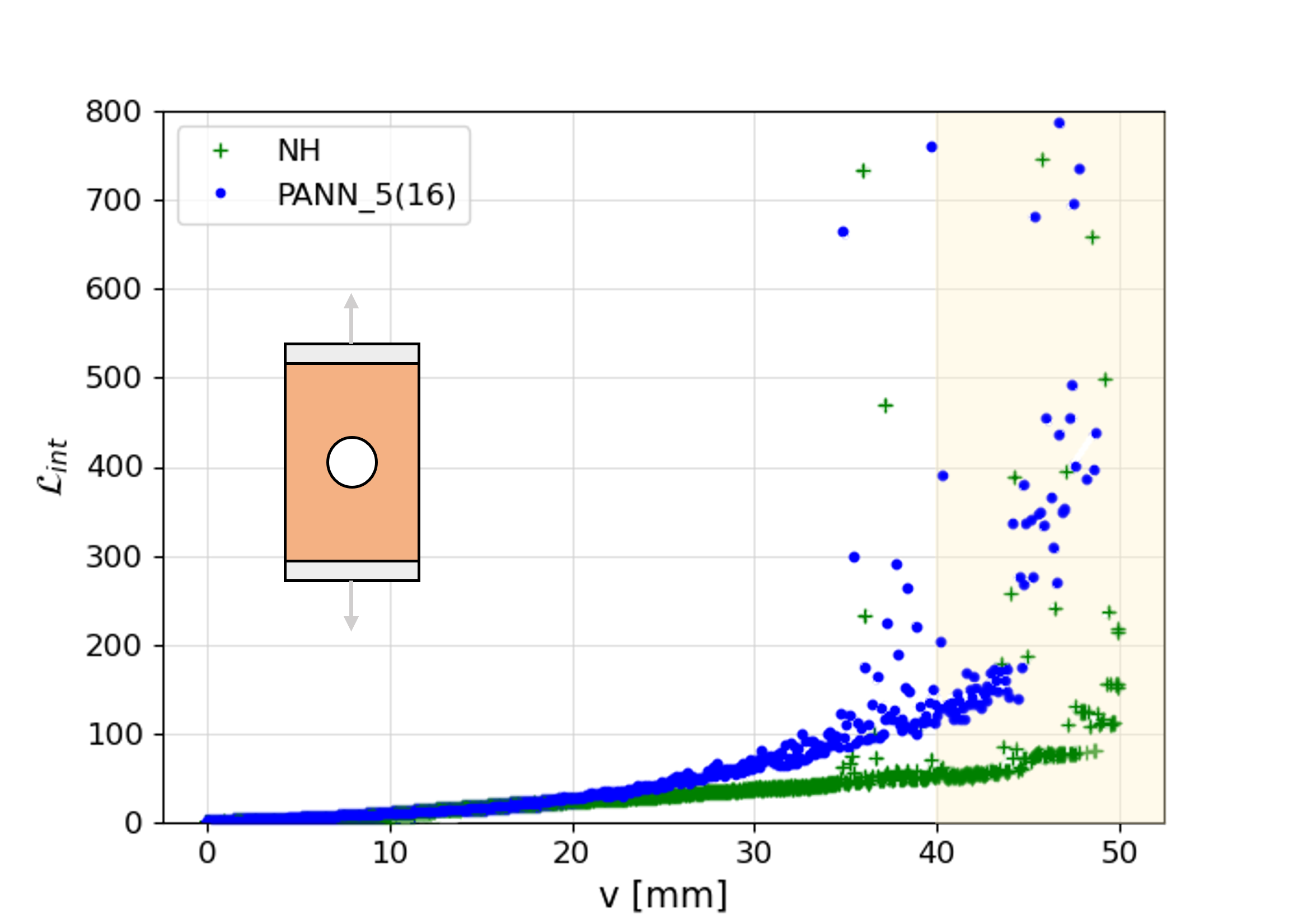  }}
        \subfigure[\revision{Experiment - ``No Hole''}]{\includegraphics[width=0.45\textwidth]{  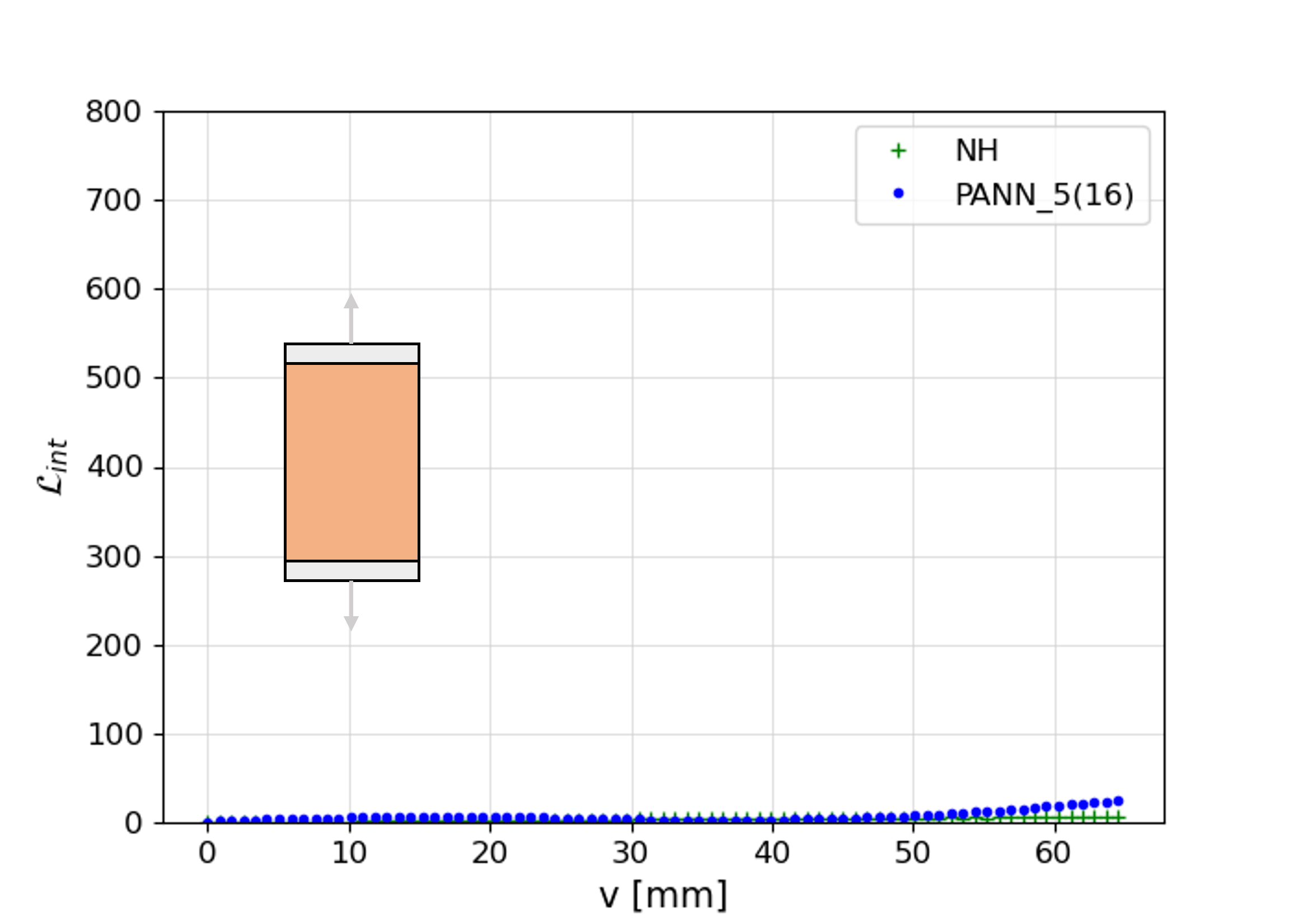  }}
	\caption{Internal metric, $\mathcal{L}_{\text{int}}$, evaluated on the test sets for the different models: PANN\_5(16) and NH. \revision{The yellow area corresponds to loadings beyond the maximal training load.}}
	\label{fig:lossesint} 
\end{figure}

\begin{figure}[t]
	\centering
        \subfigure[\revision{Experiment - ``With Hole''}]{\includegraphics[width=0.45\textwidth]{  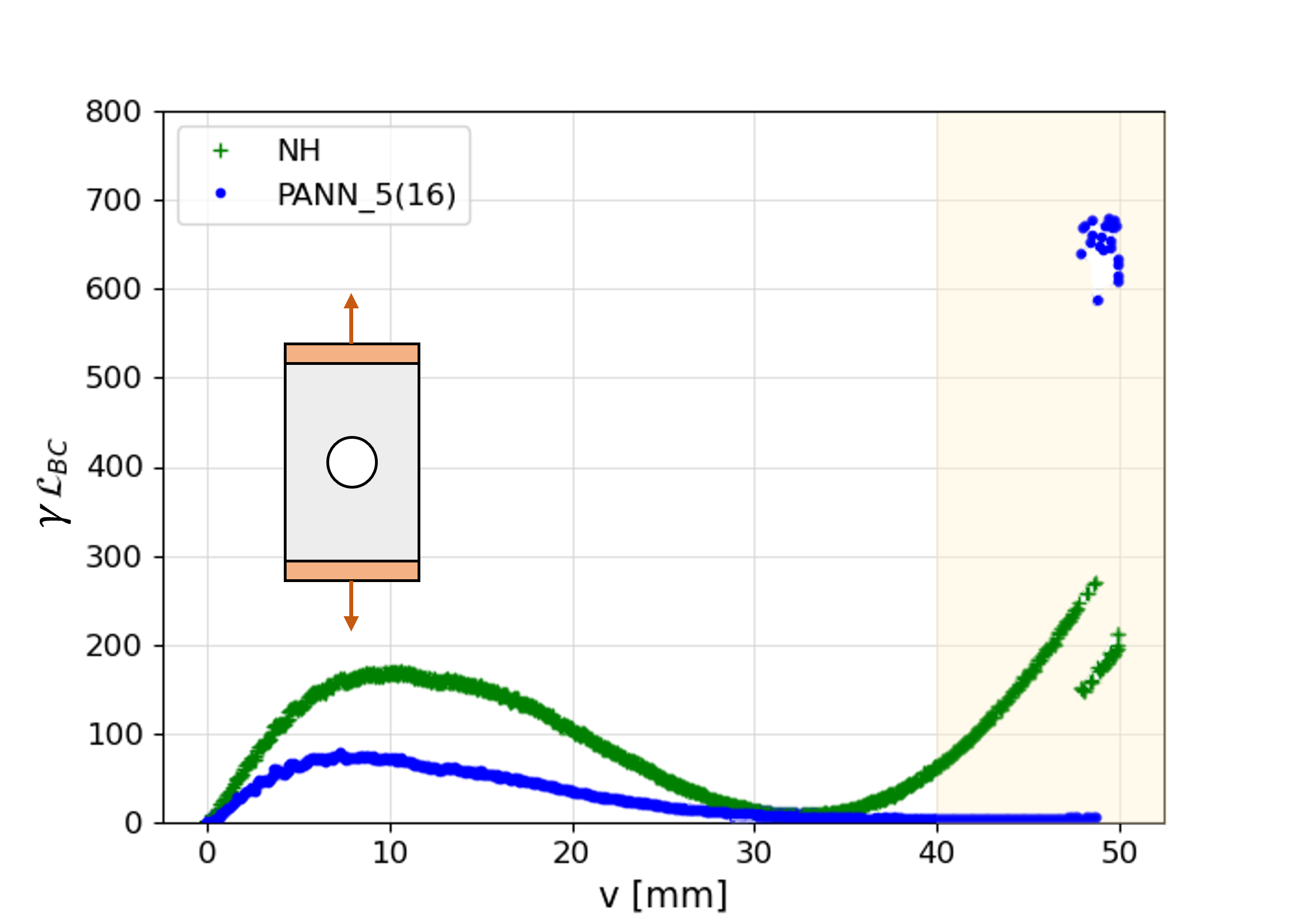  }}
        \subfigure[\revision{Experiment - ``No Hole''}]{\includegraphics[width=0.45\textwidth]{  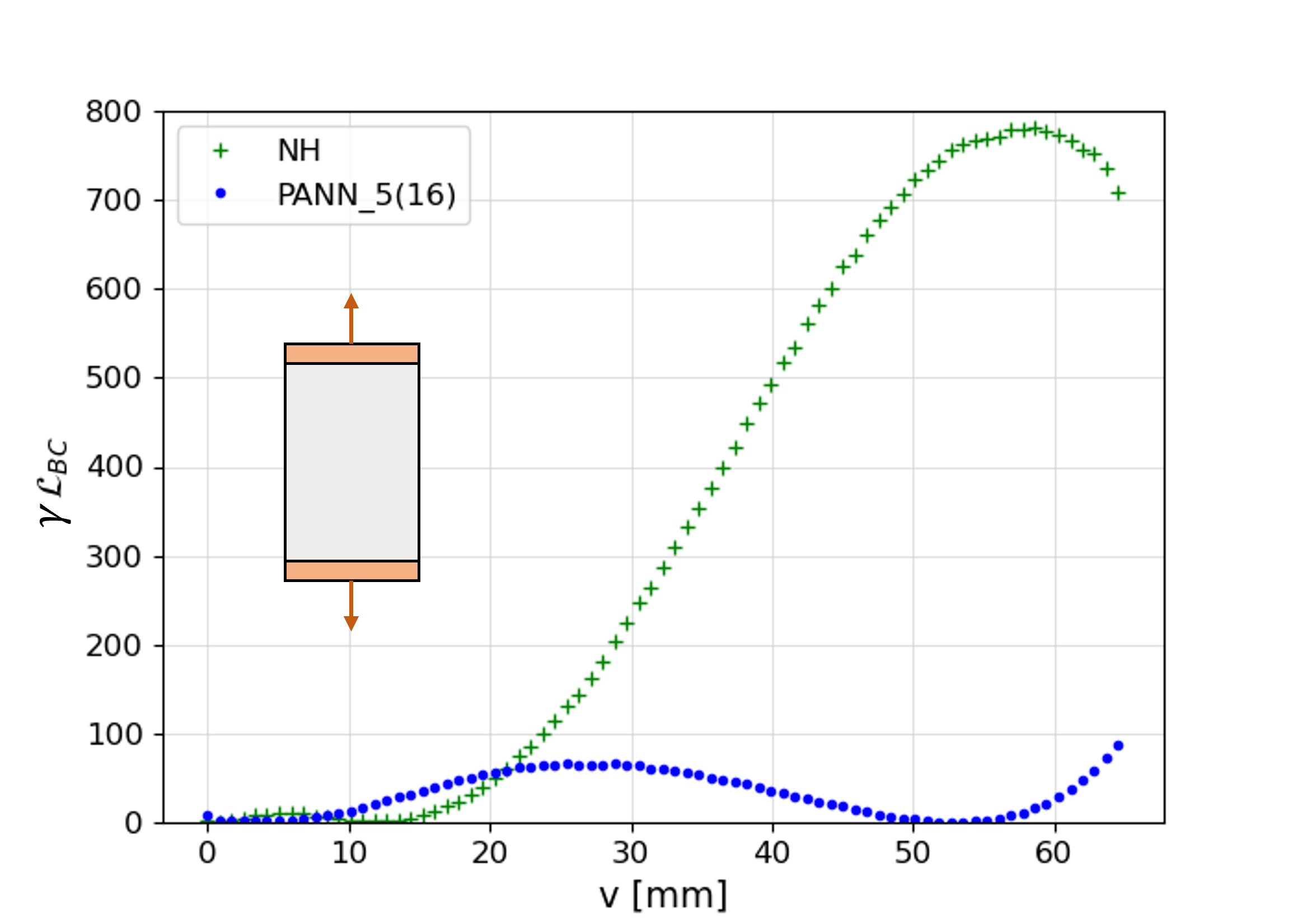  }}
	\caption{Boundary condition metric, $\gamma \mathcal{L}_{\text{BC}}$, evaluated on the test sets for the different models: PANN\_5(16) and NH. \revision{The yellow background corresponds to loadings beyond the maximal training load.}}
	\label{fig:lossesBC} 
\end{figure}

The equilibrium metrics at every loading step of the test sets for both models are presented in Figure~\ref{fig:lossesint} and \ref{fig:lossesBC},  respectively, for the internal metrics and boundary condition metric. First, the metrics evolve smoothly, without significant discontinuity between the steps close to the trained ones. This suggests that the model does not completely overfit the trained data and is able to generalize the behavior.
At the end of \revision{experiment - ``With Hole''}, some steps show discontinuous high values for both models when the loading is important. As the NH model is also impacted, the reason comes from the input deformation fields. Inaccurate DIC measurements pollute those steps and may correspond to outlier deformation element values that can be visible in Figure~\ref{fig:inv2}. The first outliers in the inner part appear at around 35~mm and at the boundary condition at around 50~mm. \revision{After the identification of the inaccurate (or converged) nodes from the DIC results, the associated outlier elements could be removed, considering a new nodal chain defining the boundary conditions and a smaller mesh for the EUCLID than the DIC.}

For \revision{experiment - ``No Hole''}, the internal loss is low for both models (under 50). Most of the total loss is represented by the boundary condition part. 
For \revision{experiment - ``With Hole''}, the internal part increases with the loading. After approximately 31.5 mm (thus after the training loads), the PANN internal error becomes higher than the NH one. This may highlight an issue in extrapolating after the trained steps. However, the boundary condition loss part shows lower values for the PANN model compared to the NH model. Finally, the last steps (around 50 mm) show important boundary condition metrics value. It may also be due to inaccurate DIC measurement on nodes at the boundary condition, as it happens for both the PANN and NH models. 

\revision{
Considering the experiment - ``No Hole'', the reaction force-displacement curves can be plotted with both models. In Figure~\ref{fig:Fv}, the black line represents the measured forces, and the green and blue markers are the NH and PANN models, respectively. This plot gives similar information to the BC loss in Figure~~\ref{fig:lossesBC}(a). The R2 score value for the test set of the experiment - ``With Hole'' is  0.95, and 0.97 for the NH model. The score computed on the experiment - ``No Hole'' is 0.98 for the PANN, and 0.86 for the NH model.
\begin{figure}[t]
	\centering
        \subfigure[Experiment - ``With Hole'']{\includegraphics[width=0.45\textwidth]{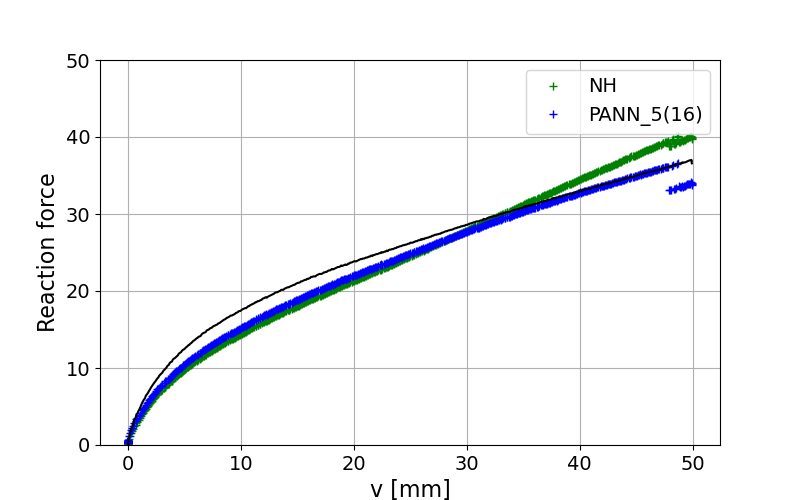}}
        \subfigure[Experiment - ``No Hole'']{\includegraphics[width=0.45\textwidth]{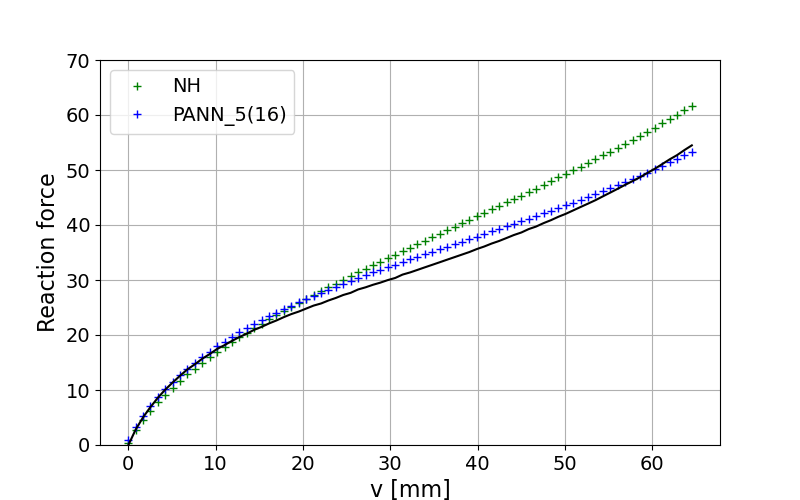}}
	\caption{\revision{Global force-displacement curves with the ground truth (black curve), the NH and PANN models.}}
	\label{fig:Fv} 
\end{figure}
}

\section{DISCUSSION}

\paragraph{Model selection}
The current study evaluates a PANN constitutive model in learning the hyperelastic behavior of a real experiment. Without knowing the true material behavior, a specific model architecture optimization has to be performed. 

First, the dataset was split into training, validation, and test parts. One experiment was selected exclusively for the test to evaluate the model's capacity to generalize on completely unseen data and sample geometry. 
Different model architectures have been evaluated and compared through the use of the equilibrium loss on the validation set. The different tested architectures presented a large range of model sizes, from small architectures: PANN\_2(8) with only 145 trainable parameters up to PANN\_5(64) with 17,217 trainable parameters.
It can be seen that architecture plays an important role in learning, as highlighted by very different efficiencies (almost a factor of two between the best and worst tested models).
Adding more layers and neurons does not seem to improve the model yet does not result in a specific overfit (at least monitored by the chosen validation steps). Noise sensitivity may play an important role when the model is composed of too many parameters~\cite{roux2024comprehensive}. From real or synthetic data, a specific focus on the optimal network architecture optimization~\cite{rosenkranz2023comparative,tacc2023benchmarking}, is an exciting perspective~\cite{he2021automl}. 
The model PANN\_5(16) was selected as a reference model. 
Trained 5 times, the final loss results presented a standard deviation of 2.8.

From an initial pre-evaluation, the chosen PANN\_5(16) model is well adapted to mimic the NH behavior when trained with unnoisy data generated with the NH model. This shows that the PANN model has the capacity to learn the NH law and may have the agility to adapt to more complex behavior. As in the real training, the PANN did not converge to the exact NH model, it shows that it learned a different behavior. \revision{The pre-training on the NH behavior was not considered and all models were retrained from scratch to avoid the risk that the  fine-tuning did not fall the solution in a specific local minimum.}

Finally, the influence of the number and richness of training data was evaluated with the training of the model from different dataset sizes. When the model is trained with too little data ($<10$), it learns the training data but fails to generalize the behavior on the validation set. This is an example of overfitting, and it highlights the necessity of having a separate validation dataset for the monitoring of the performances. With more training data ($>20$), the model performances do not improve, and overfitting is reduced. The mechanical content in the 15-20 selected steps is enough to generalize on the validation data. Selecting the training and validation data is an important topic that was treated here with linear force sampling. An important perspective is a deeper study of the training/validation data split selection and test design~\cite{jailin2017virtual,fuhg2024stress}.

The results underscore the importance of carefully controlling the datasets used for the training, validation, and test. The mechanical content in those subsets has to be informative enough for the training to learn the constitutive model from a large range of mechanical states and challenging enough for the test to provide an accurate evaluation. Although performed with two geometries and large deformations, the two experiments used in the analysis do not exhibit complex stress/strain multi-axial states, thus limiting the potential for learning. Collecting vast amounts of data from simple mechanical tests alone is not enough; instead, emphasizing the informative mechanical content of that data is crucial for the effective training of the network.

\paragraph{Comparison with standard models}
When testing the learned behavior, three main test subsets can be identified:
\begin{itemize}
    \item Test data from \revision{experiment - ``With Hole''} for loads in between the training loads (\eg $0<R<31.5$~N). The model's efficiency on those data shows the capacity of the PANN to interpolate between probed and trained data states. 
    \item Test data from \revision{experiment - ``With Hole''} for loads beyond the training loads (\eg $R>31.5$~N). The results of the PANN in this part highlight the model's capacity to extrapolate outside the training loads yet on a similar test and sample geometry. \revision{These load ranges are presented with the yellow background in Figure~\ref{fig:lossesint} and \ref{fig:lossesBC}.}
    \item Test data from \revision{experiment - ``No Hole''}, unseen during the training. As low error can be observed, it shows the capacity of the model to generalize with different data, loadings, and geometries.
\end{itemize}

\emph{\revision{Experiment - ``No Hole''}} presents low internal error with both models. While \revision{experiment - ``With Hole''} deforms heterogeneously with the hole, \revision{experiment - ``No Hole''} is more uniform. The \revision{later} internal deformation state is quite homogeneous, leading more easily to a balanced internal solution. This internal loss is, hence, not very informative for the identification/validation of models. The interesting metric of \revision{experiment - ``No Hole''} is the boundary condition metric, showing if the behavior has been well captured for elements at the boundary.  With this loss, the PANN model outperforms the NH one for high loads. \revision{The R2 score computed on the boundary condition also shows that the PANN model leads to lower errors.}
\emph{\revision{Experiment - ``With Hole''}} highlights better results for the PANN model in the training range ($0<R<31.5$~N). For higher loads, the internal error of the NH model is lower. The internal error of the PANN increases slowly after the trained steps, it may show the beginning of inaccuracy in predicting after the trained load. 

The text draws special attention to the distinction between boundary and internal errors which, however, remain closely coupled. Notably, reducing Young's modulus to values as low as zero could minimize or completely remove internal errors at the cost of increased boundary errors. Having a lower internal NH error compared to PANN while showing an important boundary condition error may show that the material stiffness is not correctly identified by the NH model. Overall, the PANN model shows a strong capacity to learn this hyperelastic behavior with metrics that compete with and outperform traditional approaches.

When compared in the invariant space plot (Figure~\ref{fig:inv}), the invariant surface covered by \revision{experiment - ``No Hole''} is contained in the one of \revision{experiment - ``With Hole''}. It is hence expected a good model generalization on \revision{experiment - ``No Hole''}. However, as the current metrics result from the integration of the nodal force errors, it is difficult to compare the metrics of \revision{experiment - ``No Hole''} and \revision{experiment - ``With Hole''} together as the geometries and meshes are not the same. Designing a more robust loss and metrics based on noise/uncertainty quantification is an interesting perspective.

\paragraph{Limits and perspectives}
The current procedure suffers from different limitations.
First, the whole PANN procedure is constrained by a single force measurement \revision{(and simple geometry without many free boundaries)}. Including additional force sensors, with more complex and instrumented testing machines, for example~\cite{dassonville2020toward}, will provide richer data collection and a better multiaxial state characterization. 

Second, it is essential to collect accurate measurements from the DIC. In the presented experiment, the deformation was very high (up to 200\% for \revision{experiment - ``With Hole''}), thus representing a real challenge for measurement procedures. As in the geometry of \revision{experiment - ``With Hole''}, large deformations are localized around the hole, corresponding to high DIC uncertainties. Particular attention was paid to positioning the mesh pixel-wise on the sample edge. Yet, as the sample and the background are not very contrasted, an error in the constitutive model may come from an incorrect mesh positioning. Paint sprayed on the surface may also degrade at this deformation range and introduce DIC errors. \revision{A perspective on the measurement side is also to use 3D measurements (DVC~\cite{buljac2018digital} that would give access to richer training data (and depth deformation), and avoid using plane stress assumptions.}
As already identified by pioneer papers~\cite{thakolkaran2022nn}, measurement noise impacts both the inference and, more importantly, the training procedure. Noise also affects the model selection. Compensating the noise bias in the evaluation loss is an important aspect. Training the PANN with more developed losses and less sensitivity to noise, with experimental data (such as mCRE, for example~\cite{benadynnmcre}), is a clear opportunity. 

The use of PANN as a constitutive law in finite element simulations may require special attention to other metrics, such as a non-zero tangent operator allowing to avoid vanishing gradients during the finite element convergence. This aspect may necessitate further investigations on the PANN architecture side (\eg choice of activation functions), as well as specific research on optimal learning data spaces.

One could argue that a more sophisticated hyperelastic model would probably better match the experimental behavior. However, the goal of this paper was not to perfectly identify the traditional model of the tested material but to train an AI model and appreciate the results compared to a standard approach. The results are promising as they show that an AI model can be trained and can generalize on new loading conditions. 

Finally, training the PANN using the NN-EUCLID framework is a long process. The current optimization was not adapted to each model architecture but was selected once to ensure weight convergence. \revision{As such, no specific convergence speed tendency was observed for the different model architectures (with respect to the depth and number of trainable parameters).} Adapting the learning rate to each model size and applying pre-training strategies may speed up the procedure. With faster convergence time, finer hyperparameter optimization could be carried out.

\section{CONCLUSION}

This study presents the application of a thermodynamics-augmented neural network learning real hyperelastic behavior. Through the use of two uni-axial experiments instrumented by digital image correlation and force sensors, the PANN model was trained within an EUCLID framework and validated. The experiments, which achieved axial deformations of over 200\% and showcased marked non-linear behavior, provided robust training, validation, and test datasets. 26 loading steps from one experiment were employed for training and validation, while an extensive set of 635 loading steps from both experiments was used for the test. Dealing with real material data, the true behavior was not accessible, so a first model architecture selection was proposed.
Comparative analysis of our model with different metrics highlighted the PANN approach's capability in modeling material behavior, particularly when extrapolating beyond the training load amplitudes, indicating the model's robustness and predictive accuracy.

The scope of the present paper is limited to hyperelastic behavior. Yet, a growing trend is to represent also history-dependent material behavior with neural networks \cite{wu2020recurrent, zhang2020using, gorji2020potential, he2022thermodynamically, benady2024unsupervised, fuhg2023modular} and training such kind of networks with experimental data remains today a challenge.

\section*{Acknowledgements}

This project was made possible by the \textit{Agence Nationale de la Recherche} (ANR) grant No. ANR-22-CPJ2-0046-01.
\\
Part of this project has received funding from the European Research Council (ERC) under the European Union's Horizon 2020 research and innovation program (grant agreement No. 101002857).

\section*{Ethical Statement/Conflict of Interest}
The authors have no conflict of interest to declare.

\bibliographystyle{unsrt} 
\bibliography{Biblio} 

\begin{thebibliography}{10}

\bibitem{dornheim2024neural}
Johannes Dornheim, Lukas Morand, Hemanth~Janarthanam Nallani, and Dirk Helm.
\newblock Neural networks for constitutive modeling: From universal function
  approximators to advanced models and the integration of physics.
\newblock {\em Archives of Computational Methods in Engineering},
  31(2):1097--1127, 2024.

\bibitem{neggers2018big}
Jan Neggers, Olivier Allix, Fran{\c{c}}ois Hild, and St{\'e}phane Roux.
\newblock Big data in experimental mechanics and model order reduction:
  today’s challenges and tomorrow’s opportunities.
\newblock {\em Archives of Computational Methods in Engineering}, 25:143--164,
  2018.

\bibitem{roux2020optimal}
St{\'e}phane Roux and Fran{\c{c}}ois Hild.
\newblock Optimal procedure for the identification of constitutive parameters
  from experimentally measured displacement fields.
\newblock {\em International Journal of Solids and Structures}, 184:14--23,
  2020.

\bibitem{Herrmann2024DeepLI}
Leon~Alexander Herrmann and Stefan Kollmannsberger.
\newblock Deep learning in computational mechanics: a review.
\newblock {\em Computational Mechanics}, 2024.

\bibitem{leygue2018data}
Adrien Leygue, Michel Coret, Julien R{\'e}thor{\'e}, Laurent Stainier, and
  Erwan Verron.
\newblock Data-based derivation of material response.
\newblock {\em Computer Methods in Applied Mechanics and Engineering},
  331:184--196, 2018.

\bibitem{leygue2019non}
Adrien Leygue, Rian Seghir, Julien R{\'e}thor{\'e}, Michel Coret, Erwan Verron,
  and Laurent Stainier.
\newblock Non-parametric material state field extraction from full field
  measurements.
\newblock {\em Computational Mechanics}, 64(2):501--509, 2019.

\bibitem{stainier2019model}
Laurent Stainier, Adrien Leygue, and Michael Ortiz.
\newblock Model-free data-driven methods in mechanics: material data
  identification and solvers.
\newblock {\em Computational Mechanics}, 64(2):381--393, 2019.

\bibitem{raissi2019physics}
Maziar Raissi, Paris Perdikaris, and George~E Karniadakis.
\newblock {Physics-informed neural networks: A deep learning framework for
  solving forward and inverse problems involving nonlinear partial differential
  equations}.
\newblock {\em Journal of Computational physics}, 378:686--707, 2019.

\bibitem{abueidda2023enhanced}
Diab~W Abueidda, Seid Koric, Erman Guleryuz, and Nahil~A Sobh.
\newblock Enhanced physics-informed neural networks for hyperelasticity.
\newblock {\em International Journal for Numerical Methods in Engineering},
  124(7):1585--1601, 2023.

\bibitem{masi2022multiscale}
Filippo Masi and Ioannis Stefanou.
\newblock {Multiscale modeling of inelastic materials with Thermodynamics-based
  Artificial Neural Networks (TANN)}.
\newblock {\em Computer Methods in Applied Mechanics and Engineering},
  398:115190, 2022.

\bibitem{fernandez2021anisotropic}
M~Fern{\'a}ndez, M~Jamshidian, T~B{\"o}hlke, K~Kersting, and O~Weeger.
\newblock Anisotropic hyperelastic constitutive models for finite deformations
  combining material theory and data-driven approaches with application to
  cubic lattice metamaterials.
\newblock {\em Computational Mechanics}, 67(2):653--677, 2021.

\bibitem{klein2022polyconvex}
Dominik~K Klein, Mauricio Fern{\'a}ndez, Robert~J Martin, Patrizio Neff, and
  Oliver Weeger.
\newblock Polyconvex anisotropic hyperelasticity with neural networks.
\newblock {\em Journal of the Mechanics and Physics of Solids}, 159:104703,
  2022.

\bibitem{as2022mechanics}
Faisal As'~ad, Philip Avery, and Charbel Farhat.
\newblock A mechanics-informed artificial neural network approach in
  data-driven constitutive modeling.
\newblock {\em International Journal for Numerical Methods in Engineering},
  123(12):2738--2759, 2022.

\bibitem{linden2023neural}
Lennart Linden, Dominik~K Klein, Karl~A Kalina, J{\"o}rg Brummund, Oliver
  Weeger, and Markus K{\"a}stner.
\newblock Neural networks meet hyperelasticity: A guide to enforcing physics.
\newblock {\em Journal of the Mechanics and Physics of Solids}, page 105363,
  2023.

\bibitem{linka2023new}
Kevin Linka and Ellen Kuhl.
\newblock {A new family of Constitutive Artificial Neural Networks towards
  automated model discovery}.
\newblock {\em Computer Methods in Applied Mechanics and Engineering},
  403:115731, 2023.

\bibitem{amos2017input}
Brandon Amos, Lei Xu, and J~Zico Kolter.
\newblock Input convex neural networks.
\newblock In {\em International Conference on Machine Learning}, pages
  146--155. PMLR, 2017.

\bibitem{hernandez2021structure}
Quercus Hern{\'a}ndez, Alberto Bad{\'\i}as, David Gonz{\'a}lez, Francisco
  Chinesta, and El{\'\i}as Cueto.
\newblock Structure-preserving neural networks.
\newblock {\em Journal of Computational Physics}, 426:109950, 2021.

\bibitem{costecalde2023data}
L{\'e}na Costecalde, Adrien Leygue, Michel Coret, and Erwan Verron.
\newblock {Data-Driven Identification of hyperelastic models by measuring the
  strain energy density field}.
\newblock {\em Rubber Chemistry and Technology}, 96(4):443--454, 2023.

\bibitem{lucas1981iterative}
Bruce~D Lucas and Takeo Kanade.
\newblock An iterative image registration technique with an application to
  stereo vision.
\newblock In {\em IJCAI'81: 7th international joint conference on Artificial
  intelligence}, volume~2, pages 674--679, 1981.

\bibitem{sutton1983determination}
Michael~A Sutton, WJ~Wolters, WH~Peters, WF~Ranson, and SR~McNeill.
\newblock Determination of displacements using an improved digital correlation
  method.
\newblock {\em Image and vision computing}, 1(3):133--139, 1983.

\bibitem{bay2008methods}
Brian~K Bay.
\newblock Methods and applications of digital volume correlation.
\newblock {\em The Journal of Strain Analysis for Engineering Design},
  43(8):745--760, 2008.

\bibitem{buljac2018digital}
Ante Buljac, Cl{\'e}ment Jailin, Arturo Mendoza, Jan Neggers, Thibault
  Taillandier-Thomas, Amine Bouterf, Benjamin Smaniotto, Fran{\c{c}}ois Hild,
  and St{\'e}phane Roux.
\newblock Digital volume correlation: review of progress and challenges.
\newblock {\em Experimental Mechanics}, 58:661--708, 2018.

\bibitem{hild2012comparison}
Fran{\c{c}}ois Hild and St{\'e}phane Roux.
\newblock Comparison of local and global approaches to digital image
  correlation.
\newblock {\em Experimental Mechanics}, 52(9):1503--1519, 2012.

\bibitem{mathieu2015estimation}
Florent Mathieu, Hugo Leclerc, Fran{\c{c}}ois Hild, and St{\'e}phane Roux.
\newblock {Estimation of elastoplastic parameters via weighted FEMU and
  integrated-DIC}.
\newblock {\em Experimental Mechanics}, 55:105--119, 2015.

\bibitem{dassonville2020toward}
Thibault Dassonville, Martin Poncelet, and Nicolas Auffray.
\newblock Toward a homogenizing machine.
\newblock {\em International Journal of Solids and Structures}, 191:534--549,
  2020.

\bibitem{dalemat2019measuring}
Marie Dal{\'e}mat, Michel Coret, Adrien Leygue, and Erwan Verron.
\newblock Measuring stress field without constitutive equation.
\newblock {\em Mechanics of Materials}, 136:103087, 2019.

\bibitem{flaschel2021unsupervised}
Moritz Flaschel, Siddhant Kumar, and Laura De~Lorenzis.
\newblock {Unsupervised discovery of interpretable hyperelastic constitutive
  laws}.
\newblock {\em Computer Methods in Applied Mechanics and Engineering},
  381:113852, 2021.

\bibitem{li2022equilibrium}
LF~Li and Chang~Q Chen.
\newblock Equilibrium-based convolution neural networks for constitutive
  modeling of hyperelastic materials.
\newblock {\em Journal of the Mechanics and Physics of Solids}, 164:104931,
  2022.

\bibitem{benady2024unsupervised}
Antoine Benady, Emmanuel Baranger, and Ludovic Chamoin.
\newblock Unsupervised learning of history-dependent constitutive material laws
  with thermodynamically-consistent neural networks in the modified
  constitutive relation error framework.
\newblock {\em Computer Methods in Applied Mechanics and Engineering},
  425:116967, 2024.

\bibitem{benadynnmcre}
Antoine Benady, Emmanuel Baranger, and Ludovic Chamoin.
\newblock Nn-mcre: A modified constitutive relation error framework for
  unsupervised learning of nonlinear state laws with physics-augmented neural
  networks.
\newblock {\em International Journal for Numerical Methods in Engineering},
  125(8):e7439, 2024.

\bibitem{jordan2020neural}
Benoit Jordan, Maysam~B Gorji, and Dirk Mohr.
\newblock Neural network model describing the temperature-and rate-dependent
  stress-strain response of polypropylene.
\newblock {\em International Journal of Plasticity}, 135:102811, 2020.

\bibitem{diamantopoulou2021stress}
Marianna Diamantopoulou, Nikolaos Karathanasopoulos, and Dirk Mohr.
\newblock {Stress-strain response of polymers made through two-photon
  lithography: Micro-scale experiments and neural network modeling}.
\newblock {\em Additive Manufacturing}, 47:102266, 2021.

\bibitem{pierre2023discovering}
Skyler R~St Pierre, Divya Rajasekharan, Ethan~C Darwin, Kevin Linka, Marc~E
  Levenston, and Ellen Kuhl.
\newblock Discovering the mechanics of artificial and real meat.
\newblock {\em Computer Methods in Applied Mechanics and Engineering},
  415:116236, 2023.

\bibitem{flaschel2023automated}
Moritz Flaschel, Huitian Yu, Nina Reiter, Jan Hinrichsen, Silvia Budday, Paul
  Steinmann, Siddhant Kumar, and Laura De~Lorenzis.
\newblock Automated discovery of interpretable hyperelastic material models for
  human brain tissue with {EUCLID}.
\newblock {\em Journal of the Mechanics and Physics of Solids}, 180:105404,
  2023.

\bibitem{thakolkaran2022nn}
Prakash Thakolkaran, Akshay Joshi, Yiwen Zheng, Moritz Flaschel, Laura
  De~Lorenzis, and Siddhant Kumar.
\newblock {NN-EUCLID: Deep-learning hyperelasticity without stress data}.
\newblock {\em Journal of the Mechanics and Physics of Solids}, 169:105076,
  2022.

\bibitem{kingma2014adam}
Diederik~P Kingma and Jimmy Ba.
\newblock {Adam: A method for stochastic optimization}.
\newblock {\em arXiv preprint arXiv:1412.6980}, 2014.

\bibitem{sciuti2021benefits}
Vinicius~F Sciuti, Rodrigo~B Canto, Jan Neggers, and Fran{\c{c}}ois Hild.
\newblock On the benefits of correcting brightness and contrast in global
  digital image correlation: Monitoring cracks during curing and drying of a
  refractory castable.
\newblock {\em Optics and Lasers in Engineering}, 136:106316, 2021.

\bibitem{mendoza2019complete}
Arturo Mendoza, Jan Neggers, Fran{\c{c}}ois Hild, and St{\'e}phane Roux.
\newblock Complete mechanical regularization applied to digital image and
  volume correlation.
\newblock {\em Computer Methods in Applied Mechanics and Engineering},
  355:27--43, 2019.

\bibitem{leclerc20153}
Hugo Leclerc, Jan Neggers, Florent Mathieu, François Hild, and Stéphane Roux.
\newblock {Correli 3.0 [IDDN. FR. 001.520008. 000. SP 2015.000. 31500]}.
\newblock {\em Agence pour la Protection des Programmes, Paris (France)}, 2015.

\bibitem{geuzaine2009}
Christophe Geuzaine and Jean-Fran{\c{c}}ois Remacle.
\newblock {Gmsh: A 3-D finite element mesh generator with built-in pre-and
  post-processing facilities}.
\newblock {\em International journal for numerical methods in engineering},
  79(11):1309--1331, 2009.

\bibitem{roux2024comprehensive}
St{\'e}phane Roux and Fran{\c{c}}ois Hild.
\newblock {Comprehensive full-field measurements via Digital Image
  Correlation}.
\newblock {\em Comprehensive Mechanics of Materials}, 2024.

\bibitem{rosenkranz2023comparative}
Max Rosenkranz, Karl~A Kalina, J{\"o}rg Brummund, and Markus K{\"a}stner.
\newblock A comparative study on different neural network architectures to
  model inelasticity.
\newblock {\em International Journal for Numerical Methods in Engineering},
  124(21):4802--4840, 2023.

\bibitem{tacc2023benchmarking}
Vahidullah Ta{\c{c}}, Kevin Linka, Francisco Sahli-Costabal, Ellen Kuhl, and
  Adrian~Buganza Tepole.
\newblock Benchmarking physics-informed frameworks for data-driven
  hyperelasticity.
\newblock {\em Computational Mechanics}, pages 1--17, 2023.

\bibitem{he2021automl}
Xin He, Kaiyong Zhao, and Xiaowen Chu.
\newblock {AutoML: A survey of the state-of-the-art}.
\newblock {\em Knowledge-Based Systems}, 212:106622, 2021.

\bibitem{jailin2017virtual}
Cl{\'e}ment Jailin, Andreea Carpiuc, Kyrylo Kazymyrenko, Martin Poncelet, Hugo
  Leclerc, Fran{\c{c}}ois Hild, and St{\'e}phane Roux.
\newblock Virtual hybrid test control of sinuous crack.
\newblock {\em Journal of the Mechanics and Physics of Solids}, 102:239--256,
  2017.

\bibitem{fuhg2024stress}
Jan Fuhg, Nikolaos Bouklas, and Reese Jones.
\newblock Stress representations for tensor basis neural networks: alternative
  formulations to finger-rivlin-ericksen.
\newblock {\em Journal of Computing and Information Science in Engineering},
  pages 1--39, 2024.

\bibitem{wu2020recurrent}
Ling Wu, Van~Dung Nguyen, Nanda~Gopala Kilingar, and Ludovic Noels.
\newblock A recurrent neural network-accelerated multi-scale model for
  elasto-plastic heterogeneous materials subjected to random cyclic and
  non-proportional loading paths.
\newblock {\em Computer Methods in Applied Mechanics and Engineering}, 2020.

\bibitem{zhang2020using}
Annan Zhang and Dirk Mohr.
\newblock Using neural networks to represent von mises plasticity with
  isotropic hardening.
\newblock {\em International Journal of Plasticity}, 2020.

\bibitem{gorji2020potential}
Maysam~B Gorji, Mojtaba Mozaffar, Julian~N Heidenreich, Jian Cao, and Dirk
  Mohr.
\newblock On the potential of recurrent neural networks for modeling path
  dependent plasticity.
\newblock {\em Journal of the Mechanics and Physics of Solids}, 143:103972,
  2020.

\bibitem{he2022thermodynamically}
Xiaolong He and Jiun-Shyan Chen.
\newblock Thermodynamically consistent machine-learned internal state variable
  approach for data-driven modeling of path-dependent materials.
\newblock {\em Computer Methods in Applied Mechanics and Engineering}, 2022.

\bibitem{fuhg2023modular}
Jan~Niklas Fuhg, Craig~M. Hamel, Kyle Johnson, Reese Jones, and Nikolaos
  Bouklas.
\newblock Modular machine learning-based elastoplasticity: Generalization in
  the context of limited data.
\newblock {\em Computer Methods in Applied Mechanics and Engineering},
  407:115930, 2023.

\end{thebibliography}

\end{document}